\documentclass[aps,pra,twocolumn,floatfix,longbibliography, superscriptaddress]{revtex4-1}

\usepackage{pifont}
\usepackage[dvipsnames]{xcolor}
\definecolor{bananayellow}{rgb}{1.0, 0.88, 0.21}
\definecolor{amethyst}{rgb}{0.6, 0.4, 0.8}
\definecolor{ao(english)}{rgb}{0.0, 0.5, 0.0}

\usepackage{float}
\usepackage{epsfig}
\usepackage{bm}
\usepackage[matrix,frame,arrow]{xy}
\usepackage[applemac]{inputenc}
\usepackage[T1]{fontenc}
\usepackage{lmodern}
\usepackage[english]{babel}
\usepackage{amsmath}
\usepackage{ae}
\usepackage{amssymb}
\usepackage{color}
\usepackage{graphicx}
\usepackage{bbm}
\usepackage{url}
\usepackage{nomencl}
\usepackage{subfigure}
\usepackage{slashed}

\usepackage{fixmath}
\usepackage{booktabs}

\newcommand{\ket}[1]{|#1\rangle}
\newcommand{\braket}[2]{\langle #1|#2\rangle}
\newcommand{\ketbra}[2]{\left| #1 \rangle \langle #2 \right|}
\newcommand{\brakket}[3]{\left\langle #1\left| #2 \right| #3\right\rangle}

\newcommand{\nn}{\nonumber}

\newcommand{\figref}[1]{\mbox{Fig.~\ref{#1}}}

\renewcommand{\eqref}[1]{\mbox{Eq.~(\ref{#1})}}

\newcommand{\be}{\begin{equation}}
\newcommand{\ee}{\end{equation}}
\newcommand{\bea}{\begin{eqnarray}}
\newcommand{\eea}{\end{eqnarray}}

\usepackage{xr}
\usepackage[colorlinks]{hyperref}
\hypersetup{%
    plainpages=true,
    breaklinks=true,
    hypertexnames=false,
    pageanchor=true,
    colorlinks=true,
    linkcolor={blue},
    citecolor={red},
    urlcolor={blue},
    anchorcolor={black}
}

\newcommand{\beq}{\begin{eqnarray}}
\newcommand{\eeq}{\end{eqnarray}}

\begin{document}
	
	
\title{Interaction of Mechanical Oscillators \\ Mediated  by  the Exchange of Virtual Photon Pairs}
	
	
	\author{Omar Di Stefano}
	\affiliation{Theoretical Quantum Physics Laboratory, RIKEN Cluster for Pioneering Research, Wako-shi, Saitama 351-0198, Japan}
	\author{Alessio Settineri}
	\affiliation{Dipartimento di Scienze Matematiche e Informatiche, Scienze Fisiche e  Scienze della Terra, Universit\`{a} di Messina, I-98166 Messina, Italy}
		\author{Vincenzo Macr\`{i}}
	\affiliation{Center for Emergent Matter Science, RIKEN, Saitama
		351-0198, Japan}
		\affiliation{Dipartimento di Scienze Matematiche e Informatiche, Scienze Fisiche e  Scienze della Terra,
		Universit\`{a} di Messina, I-98166 Messina, Italy}
	\author{Alessandro Ridolfo}
	\affiliation{Center for Emergent Matter Science, RIKEN, Saitama
		351-0198, Japan}
	
	\author{Roberto  Stassi}
	\affiliation{Center for Emergent Matter Science, RIKEN, Saitama
		351-0198, Japan}
	
	\author{Anton Frisk Kockum}
	\affiliation{Center for Emergent Matter Science, RIKEN, Saitama 351-0198, Japan}

	\author{Salvatore Savasta}
		\email[corresponding author: ]{ssavasta@unime.it}
		\affiliation{Center for Emergent Matter Science, RIKEN, Saitama
		351-0198, Japan}
	\affiliation{Dipartimento di Scienze Matematiche e Informatiche, Scienze Fisiche e  Scienze della Terra,
		Universit\`{a} di Messina, I-98166 Messina, Italy}

	\author{Franco Nori}
	\affiliation{Center for Emergent Matter Science, RIKEN, Saitama
		351-0198, Japan} \affiliation{Physics Department, The University
		of Michigan, Ann Arbor, Michigan 48109-1040, USA}


	\begin{abstract}
		Two close parallel mirrors attract due to a small force (Casimir effect) originating from the quantum vacuum fluctuations of the electromagnetic field.
		These vacuum fluctuations can also induce motional forces exerted upon one mirror when the other one moves. Here we consider an optomechanical system consisting of two vibrating mirrors constituting an optical resonator. We find that motional forces can determine noticeable coupling rates between the two spatially separated vibrating mirrors. We show that, by tuning the two mechanical oscillators into resonance, energy is exchanged between them at the quantum level. This coherent motional coupling is enabled by the exchange of virtual photon pairs, originating from the dynamical Casimir effect.
		The process proposed here shows that the electromagnetic quantum vacuum is able to transfer mechanical energy somewhat like an ordinary fluid.  
		We  show that this system can also operate as a mechanical parametric down-converter even at very weak excitations.
		These results demonstrate that vacuum-induced motional forces
		open up new possibilities for the development of optomechanical quantum technologies.
		

	\end{abstract}
	
	\pacs{ 42.50.Pq, 42.50.Ct
	}
	\maketitle
	
	
	Effective interactions able to coherently couple spatially separated qubits  \cite{Majer2007} are highly desirable  for any quantum computer architecture. 
	Efficient cavity-QED schemes, where the effective long-range interaction is mediated by the vacuum field, have been proposed \cite{Sorensen1999,Imamoglu1999, Zheng2000} and realized \cite{osnaghi2001,Majer2007,filipp2011}. In these schemes, the cavity is only virtually excited and thus the requirement on its quality factor is greatly loosened.
	Based on these interactions mediated by vacuum fluctuations, a two-qubit gate has been realized \cite{DiCarlo2009} and  two-qubit entanglement has been demonstrated \cite{Majer2007}. Creation of multi-qubit entanglement \cite{Neeley2010,Neeley2010} has also been demonstrated in circuit-QED systems. 
	Very recently, it has been shown that  the exchange of virtual photons between artificial atoms can give rise to effective interactions of multiple spatially-separated atoms \cite{Stassi2017, Zhao2017}, opening the way to vacuum nonlinear optics. Moreover, it has been shown that systems where virtual photons can be created and annihilated can be used to realize many nonlinear optical processes with qubits \cite{Kockum2017, Kockum2017b}.
	Multiparticle entanglement and quantum logic gates, via virtual vibrational excitations in an ion trap,  have also been implemented \cite{Sackett2000, Leibfried2003}. 
	A recent proposal \cite{chen2018} suggests that classical driving fields can transfer quantum fluctuations between two suspended membranes in an optomechanical cavity system.

	Given these results, one may wonder
	 whether  it is possible for spatially separated mesoscopic or macroscopic bodies to interact at a quantum level by means of the vacuum fluctuations of the electromagnetic field. 
	It is known that, owing to quantum fluctuations, the electromagnetic vacuum is able, in principle, to affect the motion of objects through it, like a complex fluid \cite{Kardar1999}. For example, it can induce dissipation and decoherence effects  on the motion of moving objects \cite{Dalvit2000,Jaekel1992a}.
	By using linear dispersion theory, it has also been shown that vacuum fluctuations can induce motional forces exerted upon one mirror when the other one moves \cite{Jaekel1992b}. 
	Here we show that two spatially separated moveable mirrors, constituting a cavity-optomechanical system, can exchange energy coherently and reversibly, by exchanging virtual photon pairs. The effects described here can be experimentally demonstrated with circuit-optomechanical systems, using ultra-high-frequency mechanical micro- or nano-resonators in the GHz spectral range \cite{OConnell2010, Rouxinol2016}. Coupling such a mechanical oscillator to a superconducting qubit, quantum control over a macroscopic mechanical system has been demonstrated \cite{OConnell2010}. 
	Our results show that the electromagnetic quantum vacuum is able to transfer mechanical energy somewhat like an ordinary fluid.  It would be as if the vibration of a string (mechanical oscillator 1) could be transferred to the membrane of a microphone (mechanical oscillator 2) in the absence of air (or any excited medium filling the gap). 
	
	We consider a system constituted by two vibrating mirrors interacting via radiation pressure [see Fig.~\ref{fig:1}(a)]. 
	Very recently, entanglement between two mechanical oscillators has been demonstrated in a similar system, where, however, the two entangled  mechanical oscillators have much lower resonance frequencies and the system is optically pumped \cite{Ockeloen2017}.
	This system can be described by a Hamiltonian that is a direct generalization to two mirrors of the Law Hamiltonian, describing the coupled mirror-field system \cite{Law1995,Butera2013,Macri2017,Sala2017,Armata2017}.
	It provides a unified description of cavity-optomechanics experiments \cite{Aspelmeyer2014} and of the dynamical Casimir effect (DCE) \cite{moore1970,Nation2012,Johansson2009,Johansson2010,Wilson2011} in a cavity with a vibrating mirror \cite{Macri2017}. It has been shown \cite{Johansson2009,Johansson2010,Wilson2011,Johansson2013,felicetti2014,stassi2015,rossatto2016} that the photon pairs generated by the DCE can be used to produce entanglement . However, in the present case, the interaction and the entaglement between two mechanical oscillators is determined by virtual photon pairs.
		Both the cavity field and the position of the mirror are treated as dynamical
	variables and a canonical quantization procedure is adopted \cite{Law1995}.  
	By considering only one mechanical mode for each mirror, with resonance frequency $\omega_{ i}$ ($i = 1,\, 2$)  and bosonic operators $\hat b^{}_i$ and $\hat b_i^\dag$, the displacement operators can be expressed as $\hat x_i = X^{(i)}_{\rm zpf} (\hat b_i^\dag + \hat b^{}_i)$, where 
	$X^{(i)}_{\rm zpf}$ is the zero-point-fluctuation amplitude of the $i$th mirror.
	The mirrors form a single-mode optical resonator with frequency $\omega_{\rm c}$ and bosonic photon operators $\hat a$ and $\hat a^\dag$. The system Hamiltonian can be written as 
	$\hat H_{\rm s} = \hat H_0 + \hat H_{\rm I}$, 
	where ($\hbar =1$) $
	\hat H_0 =  \omega_{\rm c} \hat a^\dag \hat a + \sum_i \omega_{i} \hat b_i^\dag \hat b^{}_i$
	is the unperturbed Hamiltonian. The mirror-field interaction Hamiltonian can be written as 
	$\hat H_{\rm I} = \hat V_{\rm om} + \hat V_{\rm DCE}$, where $\hat V_{\rm om} = \hat a^\dag \hat a \sum_i g_i (\hat b^{}_i + \hat b_i^\dag)$ is the standard optomechanical interaction conserving the number of photons, $\hat V_{\rm DCE} = (1/2) (\hat a^2 + \hat a^{\dag 2} ) \sum_i g_i (\hat b^{}_i + \hat b_i^\dag)$ describes the creation and annihilation of photon pairs, and $g_i$ is the optomechanical coupling rate for mirror $i$. The linear dependence of the interaction Hamiltonian on the mirror operators is a consequence of the usual small-displacement assumption \cite{Law1995}. This Hamiltonian can be directly generalized to include additional cavity modes. However, in most circuit-optomechanics experiments, the electromagnetic resonator is provided by a superconducting $LC$ circuit, which only supports a {\em single} mode.

\begin{figure}\label{fig:1}
	\centering
	\includegraphics[width=  \linewidth]{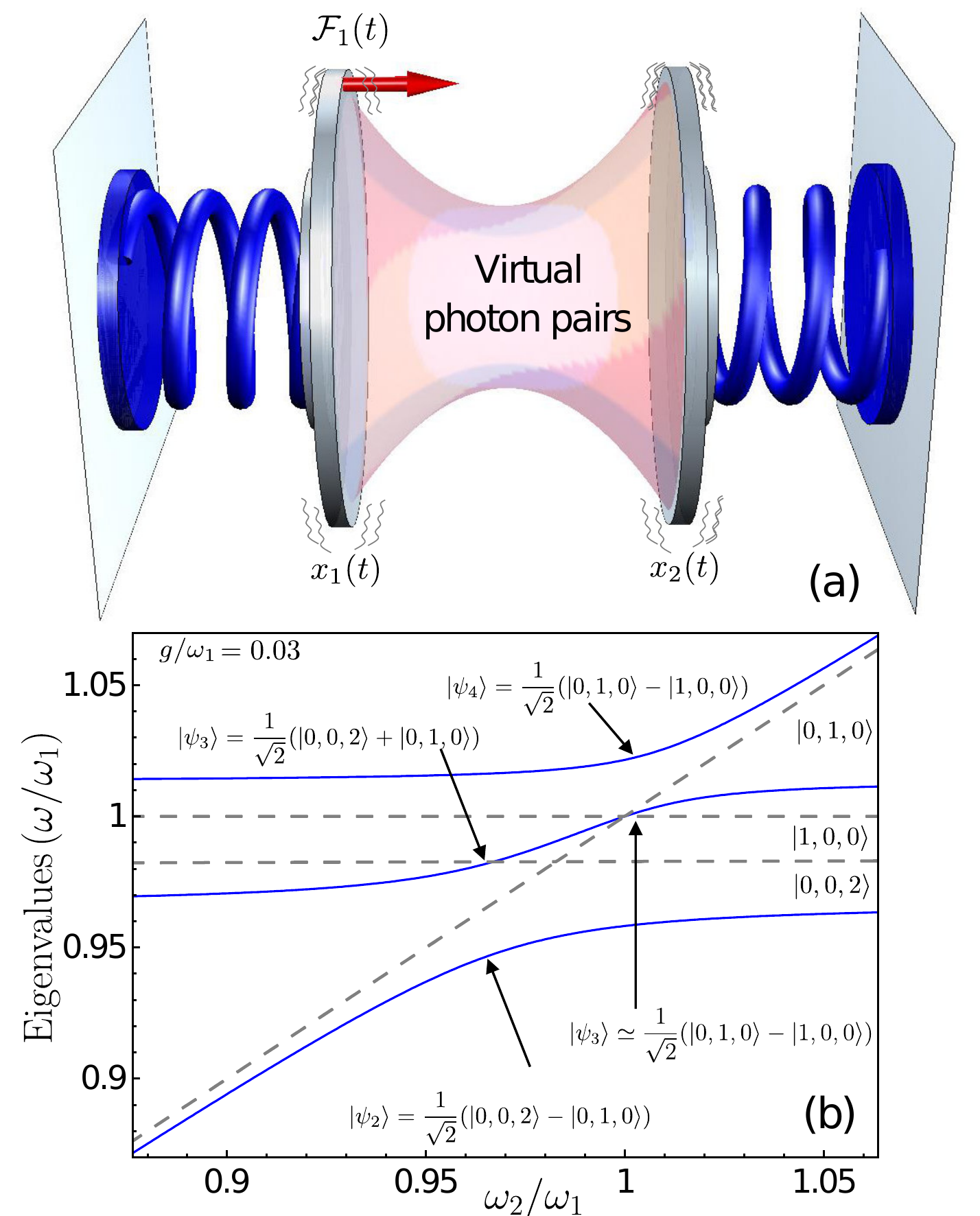}
	\caption{(a) Schematic of an optomechanical system constituted by two vibrating mirrors. If one of the two vibrating mirrors is  excited by an external drive ${\cal F}_1(t)$, its excitation can be transferred coherently and reversibly to the other mirror. The interaction is mediated by the exchange of virtual photon pairs.
		(b) Relevant energy levels of the system Hamiltonian $\hat H_s$ as a function of the ratio between the  mechanical frequency of mirror 2 and that of mirror 1. An optomechanical coupling $g/\omega_{1} = 0.03$ has been used; the cavity-mode resonance frequency is $\omega_{\rm c}  = 0.495\omega_1$. The lowest energy   anticrossing corresponds to the resonance condition for the DCE \cite{Macri2017}. The higher energy one is the signature of the mirror-mirror interaction mediated by the virtual DCE  photons.
		\label{fig:1}}
\end{figure}
	When describing most of the optomechanics experiments to date \cite{Aspelmeyer2014}, $\hat V_{\rm DCE}$ is  neglected. This is a very good approximation when
	 $\omega_i \ll \omega_{\rm c}$ (which is the most common experimental situation). However, when
	 $\omega_i$ are of the order of $\omega_{\rm c}$, $\hat V_{\rm DCE}$ cannot be neglected. We are interested in studying this regime, which can be achieved using microwave resonators and ultra-high-frequency mechanical micro- or nano-resonators \cite{OConnell2010, Rouxinol2016}. 	
	The Hamiltonian $\hat H_{\rm s}$ describes the interaction between two vibrating mirrors and the radiation pressure of a cavity field. However, the same radiation-pressure-type coupling is obtained for microwave optomechanical circuits (see, e.g., Ref.~\cite{Heikkila2014}).
	
	In order to properly describe the system dynamics, including external driving and dissipation, the coupling to external degrees of freedom needs to be considered.
    A coherent external drive of the vibrating mirror $i$ can be described by including the time-dependent Hamiltonian
    \be\label{F}
    \hat V^{}_{i}(t) = {\cal F}^{}_i(t)\, (\hat b^{}_i + \hat b_i^\dag)\, ,
    \ee
where ${\cal F}^{}_i(t)$ is equal to the external force applied to the mirror times the mechanical zero-point-fluctuation amplitude.
	Dissipation and decoherence effects are taken into account by adopting a master-equation approach. For strongly coupled hybrid quantum systems, the description offered by the standard quantum-optical master equation breaks down \cite{Beaudoin2011,Hu2015}. 
Following Refs.~\cite{Breuer2002,Ma2015, Hu2015}, we express the system-bath interaction Hamiltonian in the basis formed by the energy eigenstates of $\hat H_{\rm s}$ 
\cite{Macri2017}.

We begin our analysis by numerically diagonalizing  the Hamiltonian $\hat H_{\rm s}$ in a truncated finite-dimensional Hilbert space. The truncation is realized by only including eight Fock states for each of the three harmonic oscillators.
The blue solid curves in Fig.~\ref{fig:1}(b) describe the eigenvalue differences $E_j-E_0$ ($E_0$ is the ground-state energy) of the total Hamiltonian $\hat H_{\rm s}$ (including $\hat V_{\rm DCE}$) as a function of $\omega_2 /\omega_1$. 
For the optomechanical couplings, we use  $g_1= g_2 = g = 0.03\, \omega_{1}$. Such a coupling strength is quite high, but nevertheless below the onset of the so-called ultrastrong optomechanical coupling regime \cite{Garziano2015b, Hu2015, Macri2016}. The cavity-mode resonance frequency is fixed at $\omega_{\rm c}  = 0.495\,\omega_1$. This value is chosen close to the resonance condition for the DCE \cite{Macri2017} in order to increase the effective coupling between the mirrors.
For comparison, we also show in Fig.~\ref{fig:1}(b) (dashed grey lines) the lowest energy levels $E_{n,k_1,k_2} = \omega_{\rm c} n - \sum_i g_i^2 n^2 / \omega_{ i} + \sum_i \omega_{i} k_i$
of the standard optomechanics Hamiltonian $\hat H_{0} + \hat V_{\rm om}$. This Hamiltonian has the eigenstates
$
\ket{k_1,k_2,n} \equiv D_1(n\beta_1) \ket{k}_1 \otimes D_2(n\beta_2)\ket{k}_2 \otimes  \ket{n}_c
$,
where $|n\rangle_{\rm c}$ are the cavity Fock states and $|k \rangle_i$ are the bare mechanical states for the $i$th  mirror. 

The bare mechanical states $|k\rangle_i$ are displaced by the optomechanical interaction, $\hat D_i(n \beta_i)= \exp[{n \beta_i (\hat b_i^\dag - \hat b_i)}]$, with $\beta_i = g_i/\omega_{i}$ (see Section~\ref{Subsec:A}).
The main differences between the blue solid and the grey dashed curves are the appearance of small energy shifts, and of level anticrossings in the region $\omega_2 / \omega_1 \sim 1$. 
We indicate by $|\psi_n \rangle$ ($n = 0\,, 1\,, 2\, \dots$) the eigenvectors of $\hat H_{\rm s}$ and by  $E_n$ the corresponding eigenvalues, choosing the labelling of the states such that $E_j > E_k$ for $j>k$. The lowest energy   anticrossing corresponds to the resonance condition for the DCE \cite{Macri2017}.
The higher energy splitting in  Fig.~\ref{fig:1}(b) originates from the coherent coupling of the zero-photon states $|1,0,0 \rangle$ and $|0,1,0 \rangle$. At the minimum energy splitting $2 \lambda^{01}_{10} \simeq 2.11 \times 10^{-2} \omega_1$, the
resulting states are well approximated by $|\psi_{3,4} \rangle \simeq (1/\sqrt{2}) (|1,0,0 \rangle \pm |0,1,0 \rangle)$.
As we will show explicitly below by using perturbation theory, this {\it mirror-mirror interaction is a result of 
virtual exchange of cavity photon pairs}. When the mirrors have the same resonance frequency, an excitation in one mirror can be transferred to the other by virtually becoming a photon pair in the cavity, thanks to the DCE.
The resulting minimum energy splitting provides a measure of the effective coupling strength between the two mirrors.
At higher energy for $\omega_2 \simeq \omega_1$ a ladder of increasing level splittings, involving higher number phonon states, is present (see Section~\ref{Subsec:C}).

The origin of the higher-energy avoided-level crossing shown in Fig.~\ref{fig:1}(b) can be understood by deriving an effective Hamiltonian, using second-order perturbation theory or, equivalently, the James' method \cite{Gamel2010, Shao2017} (see Section~\ref{Subsec:B}). The resulting effective Hamiltonian, describing the coherent coupling of states 
$|1,0,0 \rangle$ and $|0,1,0 \rangle$, is
\bea\label{Heff}
\hat H_{\rm eff} =&& \Omega_1 |1,0,0 \rangle \langle 1,0,0| +  \Omega_2 |0,1,0 \rangle \langle 0,1,0|\nonumber \\ &+&(\lambda^{01}_{10}  |1,0,0 \rangle \langle 0,1,0| + {\rm H.c.})\, ,
\eea
where $\Omega_1 = \omega_1 + \Delta_{10}$ and $\Omega_2 = \omega_2 + \Delta_{01}$  denote
 the Lamb-shifted levels. The effective coupling strength is 
\be\label{lambda01}
\lambda_{10}^{01}=\sum_{k\, , q}\frac{\langle 0,1,0| \hat V_{\rm DCE} | k,q,2 \rangle \langle k,q,2| \hat V_{\rm DCE} | 1,0,0 \rangle}{E_{0,1,0}- E_{k,q,2}}\, .
\ee	
Equations~(\ref{Heff}) and (\ref{lambda01}) clearly show that the one-phonon state of mirror 1 can be transferred to mirror 2 through a virtual transition via the two-photon intermediate states $| k,q,2 \rangle$. We notice that the largest contribution is provided by the zero-phonon intermediate state ($k=q =0$). This perturbative calculation gives rise to an effective coupling strength $\lambda$ and  energy shifts $\Delta$ in good agreement with the numerical calculation shown in Fig.~\ref{fig:1}(b) (see Section~\ref{Subsec:B}). 
Analogous effective Hamiltonians can be derived for the avoided-level crossings at higher energy (see Section~\ref{Subsec:B}).

If the optomechanical couplings $g_i$ are strong enough to ensure that the DCE-induced effective coupling 
(\ref{lambda01}) becomes larger than the relevant decoherence rates in the system, the transfer of  one-phonon excitations between the two mirrors can be deterministic and reversible. Neglecting decoherence (calculations including losses can be found in Sections \ref{Subsec:E} and \ref{Subsec:F}), if the system is initially prepared in the state $|1,0,0 \rangle$, it will evolve as
\be\label{free}
|\psi(t) \rangle = \cos (\lambda_{10}^{01} t) |1,0,0 \rangle -i \sin (\lambda_{10}^{01} t) |0,1,0 \rangle\, .
\ee
After a time $t = \pi / (2\lambda_{10}^{01})$, the excitation will be completely transferred to mirror 1.
After a time $t = \pi / (4 \lambda_{10}^{01})$, the two mirrors will be in a maximally entangled motional state.

\begin{figure}
	\centering
	\includegraphics[width = 0.8 \linewidth]{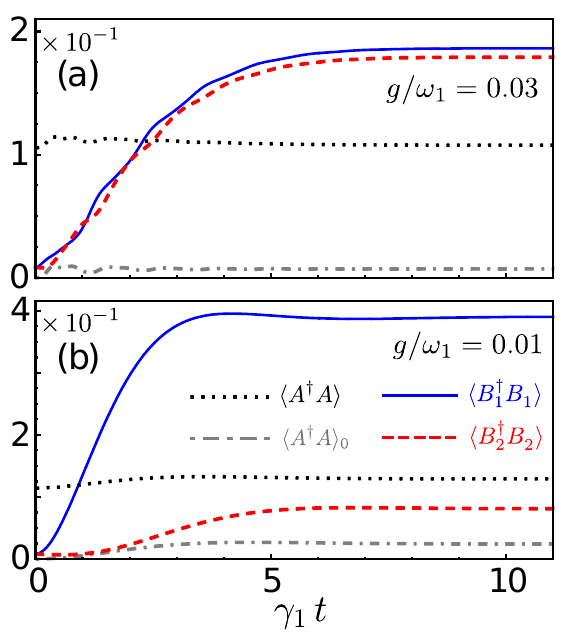}
	\caption{System dynamics for $\omega_{\rm c} \simeq 1.5 \omega_{1}$ under continuous-wave drive of  mirror 1. The blue solid and red dashed curves
		 describe   the mean phonon numbers $\langle \hat B_1^\dag \hat B_1^{} \rangle$ and $\langle \hat B_2^\dag \hat B_2^{} \rangle$, respectively, while the black dotted curve describes the mean intracavity photon number $\langle \hat A^\dag \hat A \rangle$ arising due to the DCE.
		\label{fig:2}}
\end{figure}

We now investigate the system dynamics starting from a low-temperature thermal state and introducing the excitation of mirror 1 by a single-tone continuous-wave mechanical drive
${\cal F}_1(t) = {\cal A} \cos{(\omega_{\rm d} t)}$. We numerically solve the master equation for hybrid quantum systems in a truncated Hilbert space \cite{Settineri2018}.  Figure~\ref{fig:2} shows the time evolution of the mean phonon numbers of the two mirrors $\langle \hat B_i^\dag \hat B_i \rangle$ and  the  intracavity mean photon number $\langle \hat A^\dag \hat A \rangle$. Here $\hat A, \hat B_i$ are the {\em physical} photon and phonon operators. Such operators  $\hat O = \hat A, \hat B_i$ can be defined in terms of their bare counterparts $\hat o = \hat a, \hat b_i$ as \cite{Ridolfo2012}
$
\hat O = \sum_{E_n > E_m} \langle \psi_m | (\hat o + \hat o^\dag) | \psi_n \rangle\, | \psi_m \rangle \langle \psi_n|$.
We consider the system initially in a thermal state  with a normalized thermal energy $k_B T / \omega_1 = 0.208$, corresponding to a temperature $T = 60$ mK for $\omega_1 / 2 \pi = 6$ GHz. During its time evolution, the system interacts with thermal reservoirs all with the same temperature $T$. We  use  $\gamma_1 = \gamma_2 = \gamma=  \omega_1 /260$ and $\kappa =  \gamma$ for the mechanical and photonic  loss rates. We consider a weak (${\cal A}/\gamma = 0.95$) resonant excitation of  mirror 1 ($\omega_{\rm d} = \omega_1$). We present results for two  normalized coupling strengths ($g/ \omega_{1} = 0.01,\, 0.03$), and set $\omega_2 = \omega_1$. The results shown in Fig.~\ref{fig:2}(a) demonstrate that the excitation transfer mechanism via virtual DCE photon pairs, proposed here, works very well for $g/\omega_1 = 0.03$.  In steady state, mirror 2 reaches almost the same excitation intensity as the driven mirror 1. The photon population differs only slightly from the thermal one at $t=0$, showing that a negligible amount of DCE photon pairs are generated. We also observe that the influence of temperature on the mechanical expectation values is almost negligible (see Supplemental Material). On the contrary, the cavity mode at lower frequency is much more affected by the temperature. We observe that for $g/ \omega_{1} = 0.01$, although the transfer is reduced, the effect is still measurable. The mean photon number obtained at $T=0$ is also shown for comparison (dash-dotted curves) in both the panels. The mirror-mirror excitation transfer at $g/ \omega_{1} = 0.01$ can be significantly improved (see Supplemental Material) by taking advantage of the DCE resonance condition  $\omega_c = 2 \omega_1$. However, in this case, a significant amount of real photon pairs are generated. This configuration can be used to probe the DCE effect in the presence of thermal photons.

	\begin{figure}[h!]
	\centering
	\includegraphics[width = 8.5 cm]{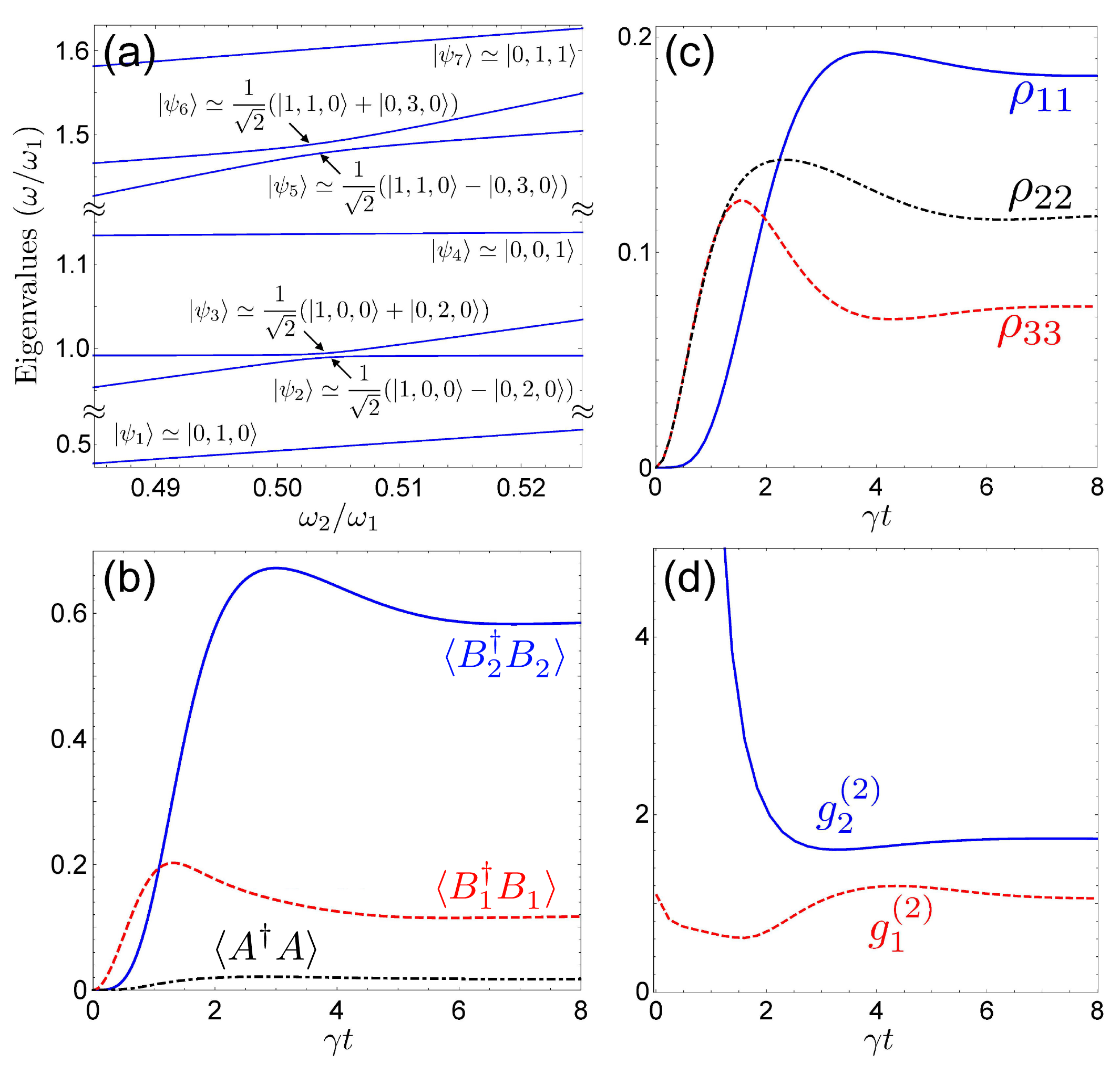} 
	\caption{Mechanical parametric down-conversion. (a) Lowest energy levels of the system Hamiltonian as a function of the ratio between the  mechanical frequency of mirror 2 and that of mirror 1. An optomechanical coupling $g/\omega_{1} = 0.12$ has been used and the cavity-mode resonance frequency is $\omega_{\rm c}  = 1.2\omega_1$. Two avoided-level crossings are clearly visible. The one at lower energy corresponds to
		the resonant coupling of the one-phonon state of mirror 1 with the two-phonon state of mirror 2, whose resonance frequency is half that of mirror 1. The higher-energy anticrossing corresponds to the resonant coupling of the states $|1,1,0 \rangle$ and $|0,3,0 \rangle$. (b) Time evolution of the mean phonon and photon numbers. (c) Time evolution of the population of the first three energy states. (d) Equal-time phonon-phonon normalized correlation functions $g^{(2)}_i(t,t)$ for the two mirrors.
		\label{fig:3}}
\end{figure}

In order to put forward the potentialities and the flexibility of this vacuum-field-mediated interaction between mechanical oscillators, we now show that this system also can  operate as a mechanical parametric down-converter. For mechanical frequencies such that  $\omega_1 \simeq 2 \omega_2$, a ladder of avoided-level crossings manifests. Two of them are shown in Fig.~\ref{fig:3}(a). 
Also in this case, the avoided-level crossings originate from the exchange of virtual photon pairs, as can be understood by using second-order perturbation theory. For example, the dominant path for the lowest energy level anticrossing goes through the intermediate state $|0,0,2 \rangle$: $|1,0,0 \rangle \leftrightarrow |0,0,2 \rangle \leftrightarrow |0,2,0 \rangle$.
We note that these avoided-level crossings, in contrast to those shown in \figref{fig:3}(a), do not conserve the excitation number.
Analogous coherent coupling effects can be observed in the ultrastrong-coupling regime of cavity QED \cite{Stassi2017,Kockum2017,Ma2015,Garziano2015,Garziano2016}.
 Using $\omega_{\rm c} = 1.2 \omega_1$ and $g/\omega_1 = 0.12$, we obtain a minimum energy	 splitting $\lambda^{02}_{10}/\omega_1 \simeq 4 \times 10^{-3}$.
We fix the resonance frequency of mirror 2 at the value providing the minimum level splitting, and calculate the system dynamics considering a weak  resonant excitation of mirror 1,
${\cal F}_1(t) = {\cal A} \cos{(\omega_{\rm d}\, t)}$,
with $\omega_{\rm d} =  (E_3+ E_2 - 2E_0)/2$, and ${\cal A}/ \gamma = 0.7$. We also used $\gamma = 2 \times 10^{-3} \omega_1$ and $\kappa = \gamma/2$. The results shown in Fig.~\ref{fig:3}(b) demonstrate a very efficient excitation transfer between the two mechanical oscillators of different frequency. We also observe that the transfer occurs even in the presence of a very weak excitation of mirror 1 (peak mean phonon number of mirror 1: $\langle \hat B_1^\dag \hat B_1 \rangle \simeq 0.2$).
It may appear  surprising that the steady-state mean phonon number of mirror 2 is significantly {\it larger} than that of mirror 1, even though it receives all the energy from the latter. This phenomenon  can be partly understood by observing that a phonon of mirror 1 converts into two phonons (each at half energy) of mirror 2. In addition, once the system decays to the state $\ket{\psi_1} \simeq \ket{0,1,0}$, the remaining excitation in mirror 2 will not be exchanged back and forth with mirror 1, since the corresponding energy level is not resonantly coupled to other energy levels [see Fig.~\ref{fig:3}(a)]. Figure~\ref{fig:3}(c) displays the populations of the three lowest-energy levels, which are the levels that are most populated at this input power. This panel confirms that $| \psi_1 \rangle$ has the higher population in steady state.

We also calculated the equal-time phonon-phonon normalized correlation functions 
\be
g_i^{(2)}(t,t) = \frac{\langle \hat B_i^\dag(t) \hat B_i^\dag(t) \hat B_i^{}(t) \hat  B_i^{}(t) \rangle}{\langle \hat  B_i^\dag(t) \hat  B_i^{}(t) \rangle^2 }\, .
\ee
The high value at early times obtained for mirror 2 [see Fig.~\ref{fig:3}(d)] confirms the {\it simultaneous excitation of phonon pairs}.

In conclusion, we demonstrated that mechanical quantum excitations can be coherently transferred among spatially-separated mechanical oscillators, through a dissipationless quantum bus, due to the exchange of virtual photon pairs. 
The experimental demonstration of these processes would show that the electromagnetic quantum vacuum is able to transfer mechanical energy somewhat like an ordinary fluid.  
The results presented here open up  exciting possibilities of applying ideas from fluid dynamics in the study of the electromagnetic quantum vacuum. 
Furthermore, these results show that the DCE in high-frequency optomechanical systems can be a versatile and powerful new resource for the development of quantum optomechanical technologies. If, in the future, it will be possible to control the interaction time (as currently realized in superconducting artificial atoms), e.g., changing rapidly the resonance frequencies of mechanical oscillators (see Section~\ref{Subsec:F}), the interaction scheme  proposed here would represent an attractive architecture for quantum information processing with optomechanical systems \cite{Stannigel2012}.
The best platform to experimentally demonstrate these results is circuit optomechanics using ultra-high-frequency ($\omega_{\rm 1}$ at 5-6 GHz) mechanical oscillators. Their quantum interaction with superconducting artificial atoms has been experimentally demonstrated \cite{OConnell2010, Rouxinol2016}. Considering instead their interaction with 
a superconducting microwave resonator should allow the observation of the effects predicted here. Specifically, combining circuit-optomechanics schemes able to increase the coupling \cite{Heikkila2014,Pirkkalainen2015} with already demonstrated ultra-high-frequency mechanical resonators \cite{OConnell2010, Rouxinol2016} represents a very promising setup for entangling spatially separated vibrations via virtual photon pairs (see Section~\ref{Subsec:G}).

\section
{\bf Supplemental Material for\\Interaction of Mechanical Oscillators \\ Mediated  by  the Exchange of Virtual Photon Pairs}


\subsection{Diagonalization of the standard optomechanics Hamiltonian} \label{Subsec:A}
We consider a system constituted by two vibrating mirrors interacting via radiation pressure [see Fig.~1(a) in the main paper].
Both the cavity field and the displacements of the mirrors are treated as dynamical variables and a canonical quantization procedure is adopted \cite{Law1995,Macri2017}.

By considering only one mechanical mode for each mirror, with resonance frequency $\omega_{
	i}$ ($i = 1,\, 2$)  and bosonic operators $\hat b_i$ and $\hat b_i^\dag$, the displacement operators can be expressed as $\hat x_i = X^{(i)}_{\rm zpf} (\hat b_i^\dag + \hat b_i)$, where 
$X^{(i)}_{\rm zpf}$ is the zero-point-fluctuation amplitude of the $i$th mirror.
We also consider a single-mode optical resonator with frequency $\omega_{\rm c}$ and bosonic photon operators $\hat a$ and $\hat a^\dag$. The system Hamiltonian can be written as 
$\hat H_{\rm s} = \hat H_0 + \hat H_{\rm I}\, ,
$
where
\begin{equation}\label{H0}
\hat H_0 =  \omega_{\rm c} \hat a^\dag \hat a +  \omega_{ 1} \hat b^\dag_1 \hat b_1 +   \omega_{2} \hat b^\dag_2 \hat b_2\, ,
\end{equation}
is the unperturbed Hamiltonian. The Hamiltonian describing the mirror-field interaction is
\begin{equation}\label{HI}
\hat H_{\rm I} = (\hat a + \hat a^\dag)^2\sum_{i=1,2} \frac{ g_i}{2} (\hat b_i + \hat b_i^\dag )\, ,
\end{equation}
where $g_i$ are the coupling rates. {Eq.~(\ref{HI}) is a direct generalization of the Law optomechanical Hamiltonian \cite{Law1995}. The linear dependence of the interaction Hamiltonian on the mirror operators is a consequence of the usual small-displacement assumption \cite{Law1995}. Once such linear dependence is assumed, the generalization (\ref{HI}) to two mirrors, coupled to the same optical resonator, is straightforward. Equation~(\ref{HI}) has a clear physical meaning: the radiation pressure force acting on the mechanical resonators is proportional to the square modulus of the electric field.}

By developing the photonic operators in normal order, and by defining 
new bosonic phonon and photon operators and a renormalized photon frequency, $\hat H_{\rm s}$ can be written 
as  	
\be\label{H}
\hat H_{\rm s}=\hat H_{\rm om} +\hat V_{\rm DCE}\, ,
\ee	
where $\hat V_{\rm DCE}$ is the DCE interaction term:
\begin{equation}
\hat V_{\rm DCE} = (\hat a^2  + \hat a^{\dag 2}) \sum_{i=1,2} \frac{ g_i}{2} (\hat b_i + \hat b_i^\dag )\, ,
\end{equation}
and $\hat H_{\rm om}$ is the standard optomechanics Hamiltonian:
\be
\hat H_{\rm om}=\hat H_0 + \hat V_{\rm om}
\ee
with
\begin{equation}
\hat V_{\rm om} =  \hat a^\dag \hat a\,\sum_{i=1,2} g_i (\hat b_i + \hat b_i^\dag )\, .
\end{equation}
$\hat H_{\rm om}$
can be easily diagonalized defining the  displacement operators for the two mirrors.
In particular, defining ($i=1,2$)
\be \label{Bdress}
\hat B_i= \hat b_i +\beta_i \hat a^\dag \hat a\,
\ee
with
$\beta_i=g_i/\omega_i$, we obtain

\bea
\hat H_{\rm om}= &&\omega_{\rm c}\left[1-\left(\frac{\beta_1^2\omega_1}{\omega_{\rm c}}+
\frac{\beta_2^2\omega_2}{\omega_{\rm c}}\right)
\hat a^\dag \hat a\right]\hat a^\dag \hat a 
\\ \nn  &&+\omega_{1} \hat B_1^\dag \hat B_1 +\omega_{2} \hat B_2^\dag \hat B_2\, .
\eea
It is possible to separate the Hilbert space spanned by the Hamiltonian eigenvectors into subspaces with a definite number of photons $n$.
The eigenstates of $\hat H_{\rm om}$ can be labelled by three indexes: the first two labelling the mechanical occupation  numbers (phonons) of the two mirrors, dressed by the presence of $n$ cavity photons while the third label describes the number $n$ of cavity photons. We use the following notation
%
\be
\ket{\psi_{k,q,n}}=\ket{k_n} \otimes\ket{q_n} \otimes  \ket{n}_c \equiv \ket{k,q,n} \,. 
\ee
In particular, the  photon occupation number $n$ determines the $n$th cavity-photon subspace, while the first two kets ($|k_n \rangle$ and $|q_n \rangle$) are the displaced mechanical Fock  states, respectively, for the first and second  mirror.
The action of the dressed phonon operators on the eigenstates satisfy 
the relations
\bea
&\hat B_1& \ket{k_n,q_n,n}=\sqrt{k}\, \ket{(k-1)_n,q_n,n}\, ,\\ \nn  &\hat B_2&\, \ket{k_n,q_n,n}=\sqrt{q}\, \ket{k_n,(q-1)_n,n}\, ,\\ \nn
& \hat B_1^\dag&\,  \ket{k_n,q_n,n}=\sqrt{(k+1)}\, \ket{(k+1)_n,q_n,n} ,\\ \nn \quad &\hat B_2^\dag&\, \ket{k_n,q_n,n}=\sqrt{(q+1)}\, \ket{k_n,(q+1)_n,n}.
\eea
The explicit expression of the single displaced Fock state $|k_n \rangle_i$ for the $i$th mirror  is (note that from \eqref{Bdress} and in the subspace with $n$ cavity photons we have $\hat B^\dag_i=\hat b^\dag_i +n\beta_i \hat I_i$)
\be \label{kn}
| k_n \rangle_i
=\frac{1}{\sqrt {k!}}\hat B_i^{\dag k}|0_n\rangle_i
=
\frac{1}{\sqrt {k!}}(\hat b^\dag_i +n\beta_i \hat I_i)^k |0_n\rangle_i 
\, ,
\ee
where $n$-photons manifold and
$|0_n\rangle_i$ is the coherent ground state for mirror $i$ with $n$ cavity photons, as is shown by the relation
\be\hat b_i |0_n\rangle_i=-n\beta_i |0_n\rangle_i \, ,\ee
obtained using \eqref{Bdress} in 
$\hat B_i |0_n\rangle_i=0$.
Using the displacement operator $\hat D(n \beta_i)= \exp[{n \beta_i ( \hat b_i-\hat b_i^\dag)}]$, we have
\be \label{0n}
|0_n\rangle_i=\hat D(n \beta_i)|0\rangle_i=\sum_j e^{-|n\beta_i|^2/2}\frac{(-n\beta_i)^j}{\sqrt{j!}}\ket{j}_i\, .\ee
In addition,  from the relation $\hat D(n \beta)\hat b^\dag \hat D^\dag(n \beta)= b^\dag +n \beta $ \cite{Scully1997}, using Eqs.~(\ref{kn}) and (\ref{0n}), we obtain
\bea
| k_n \rangle_i &=& \nn
\frac{1}{\sqrt {k!}}(\hat b^\dag_i +n\beta_i \hat I_i)^k |0_n\rangle_i =\frac{1}{\sqrt {k!}}(\hat b^\dag_i +n\beta_i \hat I_i)^k\hat D(n \beta_i)|0\rangle\\ &=&\hat D(n \beta_i)\frac{1}{\sqrt {k!}}\hat b^{\dag k}_i|0\rangle=\hat D(n \beta_i)|k_0\rangle
\eea
Finally, after a little bit of algebra, we have
\bea
&&_{i}\langle k^\prime_0| k_n^{} \rangle_i=\,
_{i}\langle k^\prime_0|[\hat D(n\beta_i)]|k_0\rangle_i=\\ \nn
&&D_{k^\prime,k}(n\beta_i)=
\sqrt{k!/k^\prime!} (n\beta_i)^{k'-k} e^{-|n\beta_i|^2/2} L_k^{k^\prime-k}(|n\beta_i|^2)\, ,
\eea
where $L_k^{p}(x)$ are the associated Laguerre polynomials.

In conclusion, the standard optomecanical Hamiltonian can be diagonalized as shown above and we obtain
\be
\hat H \ket{k,q,n}=E_{k,q,n}\ket{k,q,n}\, ,
\ee
where
\bea
E_{k,q,n}&=& \omega_{\rm c}n \nn \left[1-\left(\frac{\beta_1^2\omega_1}{\omega_{\rm c}}+
\frac{\beta_2^2\omega_2}{\omega_{\rm c}}\right)
n\right]+ \\ &+&
\omega_{1} k + \omega_{2} q\, ,
\eea
or, in more compact form [replacing for clarity the phonon labels as $(k,q)\rightarrow (k_1,k_2) $]
\be
E_{k_1,k_2,n} = \omega_{\rm c} n - \sum_i g_i^2 n^2 / \omega_{ i} + \sum_i \omega_{i} k_i.
\ee

\subsection{The DCE interaction Hamiltonian as a perturbation}
\label{Subsec:B}
In this section, we introduce the DCE interaction term. We consider this additional contribution  as a perturbation to the optomechanical Hamiltonian $\hat H_{\rm om}$. This additional term creates and destroys  photon pairs. Here we consider processes at the lowest nonzero perturbation order. Thus we limit our calculations to the subspace containing zero and two cavity photons.
The DCE interaction Hamiltonian $\hat V_{\rm DCE}$ is calculated  using second-order perturbation theory. These perturbative calculations are carried out using the James'  method \cite{ Shao2017}:

\be\label{H2}
\hat H^{(2)}_{\rm eff}= \frac{1}{i}\hat V^{I(0,2)}_{\rm DCE}(t)\int_0^t
\hat V^{I(0,2)}_{\rm DCE}(t^\prime)dt^\prime\, ,
\ee
where $$ \hat V^{I(0,2)}_{\rm DCE}(t)= e^{i\hat H t} \hat V^{(0,2)}_{\rm DCE} e^{-i\hat H t}$$ 
is the projection  operator $\hat V_{\rm DCE}$ acting in the  subspace containing $0$ and $2$ photons expressed in the interaction picture.
After some algebra, we obtain (we assume  $g_1 = g_2 \equiv g$):
\bea\label{VI}
\hat V^{I(0,2)}_{\rm DCE}(t)&&=\frac{ g}{2} \sum\limits_{\substack{k^{\phantom{\prime}}\, q^{\phantom{\prime}} \\ \nn k^\prime\,q^\prime}}
A^{k^\prime\,q^\prime}_{k^{\phantom{\prime}}\, q^{\phantom{\prime}}}
\ketbra{k_2,q_2,2}{k_0^\prime,q_0^\prime,0} e^{i\omega^{k^\prime\,q^\prime}_{k^{\phantom{\prime}}\, q^{\phantom{\prime}}}t}\\ &&+
(A^{k^\prime\,q^\prime}_{k^{\phantom{\prime}}\, q^{\phantom{\prime}}})^\dag
\ketbra{k_0^\prime,q_0^\prime,0}{k_2,q_2,2} e^{-i\omega^{k^\prime\,q^\prime}_{k^{\phantom{\prime}}\, q^{\phantom{\prime}}}t}
\eea
where 
\be
\omega^{k^\prime\,q^\prime}_{{k^{\phantom{\prime}}} q^{\phantom{\prime}}}=
2\Omega_{\rm c}+
(k^\prime-k)\omega_{1}
+(q^\prime-q)\omega_{2};
\ee
with $\Omega_{\rm c}= 1+\tilde \beta_1+\tilde \beta_2$, $\tilde\beta_i=g^2/(\omega_i \omega_{\rm c})$
.
We also have: 
$$
A^{k^\prime\,q^\prime}_{k^{\phantom{\prime}}\, q^{\phantom{\prime}}}=\brakket{k_2,q_2,2}{\hat V_{\rm DCE}}{k^\prime_0,q^\prime_0,0}\,;
$$
that can be expressed in more explicit form as

\bea
&&A^{k^\prime\,q^\prime}_{k^{\phantom{\prime}}\, q^{\phantom{\prime}}}= \nn
\sqrt{2}
\left\{ [\sqrt{k^\prime}\braket{k_2}{(k^\prime-1)_0}\right. \\ &&\left.+\sqrt{k^\prime +1}\braket{k_2}{(k^\prime+1)_0}]\braket{q_2}{q_0^\prime}\right. \\ &&\left.+ [\sqrt{q^\prime}\braket{q_2}{(q^\prime-1)_0}+\sqrt{q^\prime +1}\braket{q_2}{(q^\prime+1)_0}]\braket{k_2}{k_0^\prime} \nn
\right\}\, .
\eea
Note that 
$A^{k^\prime\,q^\prime}_{k^{\phantom{\prime}}\, q^{\phantom{\prime}}}=A_{\phantom{\dag}k^\prime\,q^\prime}^{\dag k^{\phantom{\prime}}\, q^{\phantom{\prime}}}$.
Using $D_{k',k}(2\beta_i)=\braket {k^\prime_2}{k_0}$, we have:
\bea
&&A^{k^\prime\,q^\prime}_{k^{\phantom{\prime}}\, q^{\phantom{\prime}}}= \nn
\sqrt{2}[\sqrt{k^\prime}D_{k,k^\prime-1}(2\beta_1) \\ &&+\sqrt{k^\prime +1}D_{k,k^{\prime}+1}(2\beta_1)]D_{q,q^{\prime}}(2\beta_2)+\\
&&\sqrt{2}[\sqrt{q^{\prime}}D_{q,q^\prime-1}(2\beta_2)+\sqrt{q^\prime +1}D_{q,q^\prime+1}(2\beta_2)]D_{k,k^\prime}(2\beta_1) \nn\, ,
\eea
where the matrix elements of the displacement operators can be expressed in terms of associated Laguerre polynomials: $D_{k',k}(\alpha)= \sqrt{k!/k'!} \alpha^{k'-k} e^{-|\alpha|^2/2} L_k^{k'-k}(|\alpha|^2)$.

\subsubsection{One phonon -- zero photons subspace}

The $(1+0)$ subspace containing zero photons and one phonon excitation is spanned by the eigenvectors $\ket{1,0,0}$ and  $\ket{0,1,0}$.
At $\omega_2\sim \omega_1$, these states are degenerate in absence of the $\hat V_{\rm DCE}$ interaction. In presence of such interaction, degeneracy is removed and an avoided level crossing can be observed. This effect can be described by introducing an effective Hamiltonian. Specifically:  a) we introduce \eqref{VI} into \eqref{H2}; b) we perform the integration; c) we limit the calculations to matrix elements containing zero photons; d) we transform back to the Schr\"{o}dinger picture; e) finally, we project the result into the $(1+0)$ subspace spanned by the vectors $\ket{1,0,0}$,  $\ket{0,1,0}$. We obtain

\be
\hat H_{\rm eff}=\hat H_{\rm eff}^0+[\lambda^{10}_{01} \ketbra{0,1,0}{1,0,0}
+ {\rm H.c.}],
\ee
where

\be
\hat H_{\rm eff}^0= \Omega_1 \ketbra{1,0,0}{1,0,0}
+ \Omega_2 \ketbra{0,1,0}{0,1,0},
\ee
with $\Omega_1=\omega_1+\Delta_{10}$ and $\Omega_2=\omega_2+\Delta_{01}$, and with

\be
\Delta_{10}=-\frac{ g^2}{4}\sum_{k\,q}\frac{A_{k\,q}^{ 1\,0\dag} A_{k\,q}^{ 1\,0\phantom{\dag}}}{2\Omega_{\rm c}+(k-1)\omega_1+q\omega_2};
\ee	

\be
\Delta_{01}=-\frac{ g^2}{4}\sum_{k\,q}\frac{A_{k\,q}^{0\,1\dag } A_{k\,q}^{ 0\,1\phantom{\dag}}}{2\Omega_{\rm c}+k\omega_1+(q-1)\omega_2};
\ee

\be
\lambda^{10}_{01}=-\frac{ g^2}{4}\sum_{k\,q}\frac{A_{k\,q}^{ 0\,1\dag} A_{k\,q}^{1\,0\phantom{\dag}}}{2\Omega_{\rm c}+(k-1)\omega_1+q\omega_2}.
\ee	
\begin{figure}
	\includegraphics[width = 8 cm]{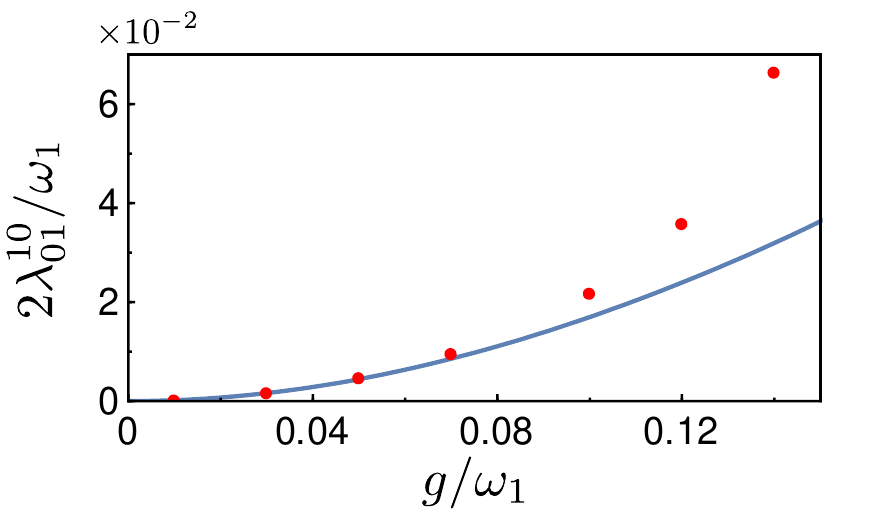}
	\caption{Comparison between the numerically calculated normalized Rabi splitting (red points)  (corresponding to twice the effective coupling between the two one-phonon states $\ket {1,0,0}$ and $\ket{0,1,0}$) and the corresponding calculation using second-order perturbation theory (solid blue curve).
	}\label{Fig2}
\end{figure}
In \figref{Fig2}, we show a comparison between the numerically calculated normalized Rabi splitting ($2\lambda_{01}^{10} \omega_1$) between the two one-phonon states $\ket {1,0,0}$ and $\ket{0,1,0}$ and the corresponding theoretical value calculated using second-order perturbation theory  
as a function of the normalized optomechanical
coupling $g/\omega_1$. The agreement is very good for $g/\omega_1$ below 0.1.

\subsubsection{Two phonons -- zero photons subspace}
The ($2+0$) subspace with zero photons in the cavity and containing two phonon excitations is spanned by the eigenvectors: $\ket{2,0,0}$, $\ket{0,2,0}$ and $\ket{1,1,0}$.
Also in this case, at $\omega_2\sim \omega_1$, these states are degenerate in the absence of the $\hat V_{\rm DCE}$ interaction. With the introduction of $\hat V_{\rm DCE}$, degeneracy is removed, and an avoided level crossing can be observed. Following the same procedure described  in the previous subsection, this effect can be described by introducing an effective Hamiltonian acting on the ($2+0$) subspace. We obtain:

\bea
\hat H_{\rm eff}&=&\hat H_{\rm eff}^0+[ \lambda_{20}^{02} \nn \ketbra{2,0,0}{0,2,0}
+\lambda_{20}^{11} \ketbra{2,0,0}{1,1,0}+ \\ &&
+\lambda_{02}^{11} \ketbra{0,2,0}{1,1,0}+
{\rm H.c.}];
\eea
where

\bea
\hat H_{\rm eff}^0&=& \Omega_{20} \ketbra{0,2,0}{0,2,0}\nn
+ \Omega_{02} \ketbra{2,0,0}{2,0,0}\\ &&+\Omega_{11}\ketbra{1,1,0}{1,1,0};
\eea
with $\Omega_{20}=2\omega_1+\Delta_{20}$, $\Omega_{11}=\omega_1+\omega_2+\Delta_{11}$  and $\Omega_{02}=2\omega_2+\Delta_{02}$, and

\be
\lambda_{20}^{02}=-\frac{ g^2}{4}\sum_{k\,q}\frac{A_{k\,q}^{ 0\,2\dag} A_{k\,q}^{2\,0\phantom{\dag}}}{2\Omega_{\rm c}+(k-2)\omega_1+q\omega_2},
\ee	

\be
\lambda_{20}^{11}=-\frac{ g^2}{4}\sum_{k\,q}\frac{A_{ k\,q}^{1\,1\dag } A_{k\,q}^{2\,0\phantom{\dag}}}{2\Omega_{\rm c}+(k-2)\omega_1+q\omega_2},
\ee	

\be
\lambda_{02}^{11}=-\frac{ g^2}{4}\sum_{k\,q}\frac{A_{ k\,q}^{1\,1\dag } A_{k\,q}^{0\,2\phantom{\dag}}}{2\Omega_{\rm c}+k\omega_1+(q-2)\omega_2},
\ee	

\be
\Delta_{20}=-\frac{ g^2}{4}\sum_{k\,q}\frac{A_{k\,q}^{ 2\,0\dag} A_{k\,q}^{ 2\,0\phantom{\dag}}}{2\Omega_{\rm c}+(k-2)\omega_1+q\omega_2},
\ee	

\be
\Delta_{02}=-\frac{ g^2}{4}\sum_{k\,q}\frac{A_{k\,q}^{0\,2\dag } A_{k\,q}^{ 0\,2\phantom{\dag}}}{2\Omega_{\rm c}+k\omega_1+(q-2)\omega_2},
\ee	

\be
\Delta_{11}=-\frac{ g^2}{4}\sum_{k\,q}\frac{A_{k\,q}^{ 1\,1\dag} A_{k\,q}^{1\,1\phantom{\dag}}}{2\Omega_{\rm c}+(k-1)\omega_1+(q-1)\omega_2}.
\ee	

A comparison of these perturbative analytical results with the numerical result is provided in the Tables I and II. The discrepancies can be ascribed to higher-order terms that at a coupling strength $g/ \omega_1 = 0.1$ provide non-negligible contributions. 


\begin{widetext}
	\begin{center}
\begin{table}[h!]
	\renewcommand\tabcolsep{10pt}
	\begin{tabular}{|*8{c|}}
		\hline
		& $2\lambda^{10}_{01}$ & $2\lambda_{20}^{11}$ & $2\lambda_{20}^{02}$ & $2\lambda^{11}_{02}$ \\
		\hline
		Numerical  $\simeq$& 0.0217& $0.0217$&0.0384&0.0167\\
		\hline
		Theoretical $\simeq$ &0.0170 &$ 0.0171$ &0.0348& 0.0177\\
		\hline
	\end{tabular}
	\caption{Comparison between the effective splittings calculated both numerically (as difference between the eigenvalues) and  analytically using the James' method \cite{ Shao2017}. 
		In particular, the theoretical  values corresponding to $2\lambda_{20}^{11}$, $2\lambda_{20}^{02}$ and $2\lambda^{11}_{02}$  are obtained by the diagonalization of a $3\times3$ matrix representing the effective Hamiltonian in the subspace with two phonon excitations and zero photons.
		The cavity-mode resonance frequency is $\omega_{\rm c}=0.85\, \omega_1$ and $\omega_2= \omega_1$.}
	\label{t1}
\end{table}
\end{center}
\end{widetext}

\begin{widetext}
	\begin{center}
\begin{table}[h!]
	\renewcommand\tabcolsep{10pt}
	\begin{tabular}{|*8{c|}}
		\hline
		&$ \Delta_{10}$  &$ \Delta_{01}$  & $\Delta_{11}$ & $\Delta_{02}$&$\Delta_{20}$ \\
		\hline
		Numerical $\simeq$ &$-0.0131$ & $-0.0159$ & $-0.0221$ & $-0.0239$ & $-0.0217$\\
		\hline
		Theoretical $\simeq$ & $-0.0120$ & $-0.0121$ & $ -0.0207$ & $-0.0199$ & $-0.0207$\\
		\hline
	\end{tabular}
	\caption{Comparison between the numerically calculated energy shifts and the analytical calculations obtained using the James' method. The mechanical frequency of mirror 2 is  $\omega_2=0.94\, \omega_1$. For this value the energy levels investigated do not interact significantly, and hence the energy shifts are not affected by the level-repulsion effect that occurs when the mirrors are on resonance with each other. The cavity-mode resonance frequency is $\omega_{\rm c}=0.85\, \omega_1$.}
	\label{t2}
\end{table}
\end{center}
\end{widetext}

\newpage

\subsection{Energy levels and splittings for different optomechanical couplings.}
\label{Subsec:C}
\begin{figure}
	\centering
	\includegraphics[width = 8 cm]{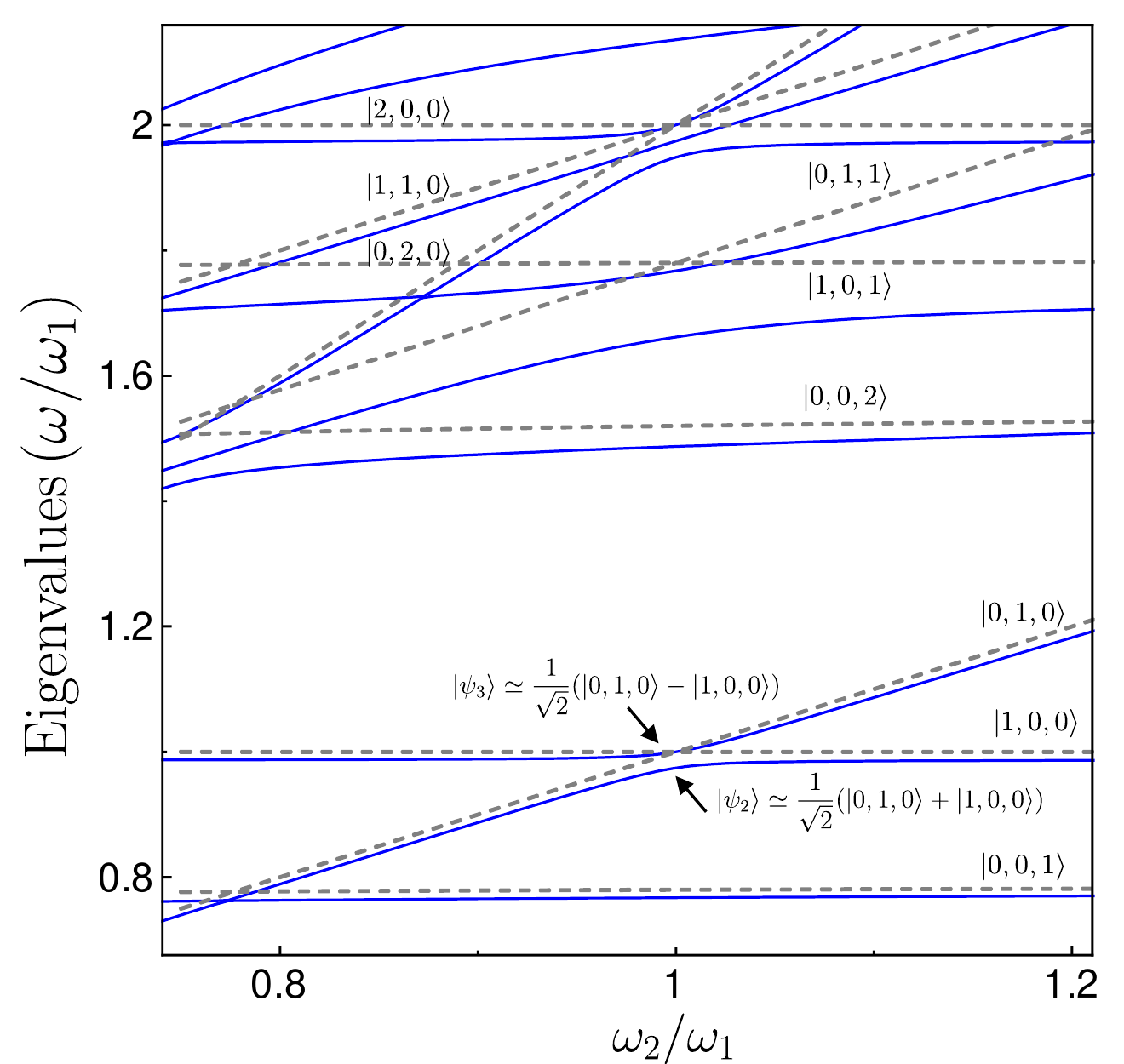}
	\caption{Lowest energy levels of the system Hamiltonian as a function of $\omega_2/\omega_1$. We used $g/\omega_1 = 0.1$ and $\omega_c/\omega_1 = 0.8$.
		\label{Sg01}}
\end{figure}
Figure~\ref{Sg01} displays the lowest energy levels $E_j - E_0$ of the system Hamiltonian
as a function of the ratio between the mechanical frequency
of mirror 2 and that of mirror 1. An optomechanical coupling
$g/\omega_1 = 0.1$ has been used, the cavity-mode resonance
frequency is $\omega_c = 0.8\, \omega_1$.
Starting 
from the lowest energy levels, we first avoided level crossing originates from the coherent coupling of the zero-photon states $|1,0,0 \rangle$ and $|0,1,0 \rangle$. At the minimum energy splitting, the
resulting states are well approximated by $|\psi_{2,3} \rangle \simeq (1/\sqrt{2}) (|1,0,0 \rangle \pm |0,1,0 \rangle)$.
As shown in the main paper and in the previous section, this  mirror-mirror interaction is a result of  virtual exchange of cavity photon pairs.
This coherent coupling is greatly enhanced by the presence of a cavity photon, resulting in the larger splitting $(E_6-E_5)$, corresponding to the states $|\psi_{5,6} \rangle \simeq (1/\sqrt{2}) (|1,0,1 \rangle \pm |0,1,1 \rangle)$. At higher energy, at $\omega_2 / \omega_1 \sim 1$, $\hat V_{\rm DCE}$
removes the degeneracy between the three states $|2,0,0 \rangle$, $|0,2,0 \rangle$, and $|1,1,0 \rangle$, determining a two-phonon coupling between the two mirrors.

\begin{figure}
	\centering
	\includegraphics[width = 8.5 cm]{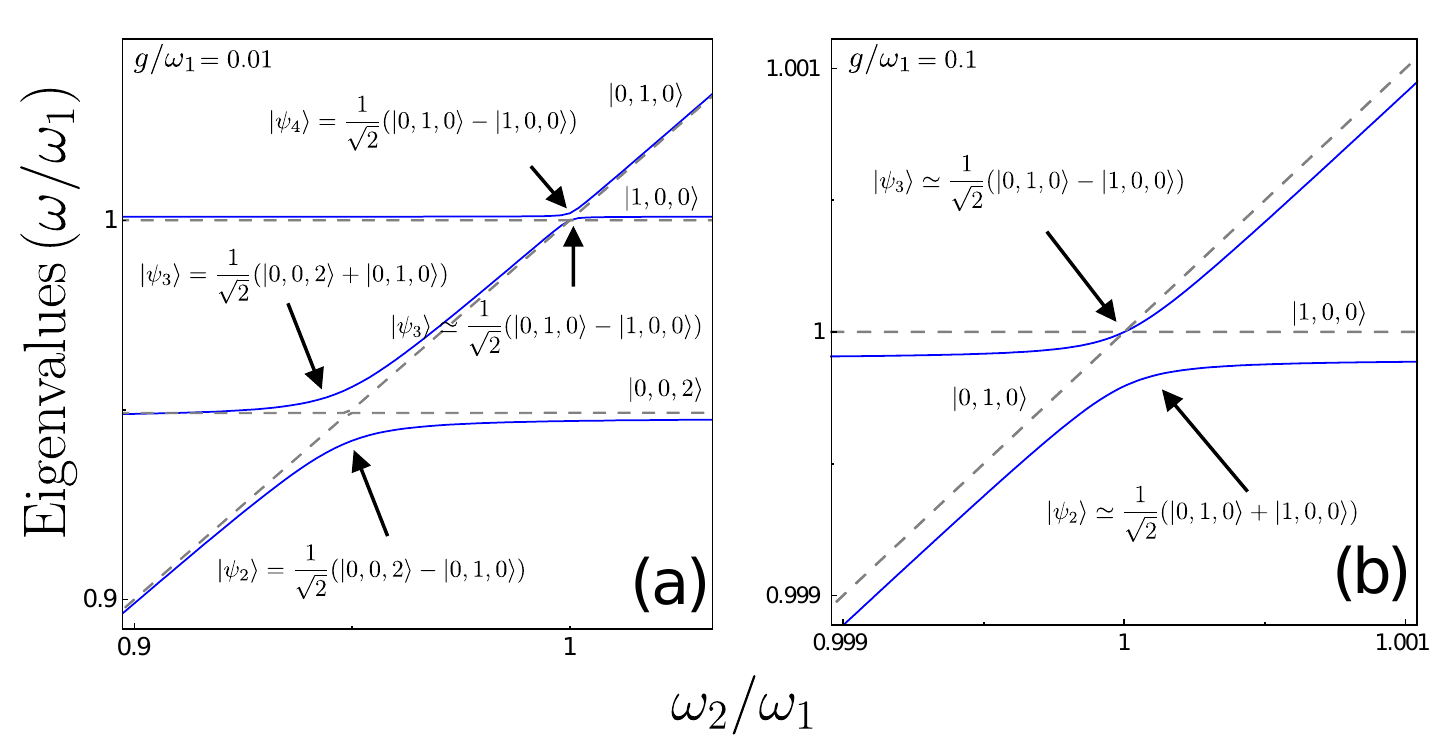}
	\caption{Relevant lowest energy levels of the system Hamiltonian as a function of $\omega_2/\omega_1$.  Panel (a) has been obtained using  $g/\omega_1 = 0.01$ and $\omega_c /\omega_1 = 0.475$. Panel (b) has been obtained with the same parameters of \figref{Sg01}.
		\label{Sg01001}}
\end{figure}

Figure~\ref{Sg01001} shows the relevant energy levels of the system Hamiltonian
$\hat H_s$ as a
function of the ratio $\omega_2/\omega_1$. 
For the  panel (a) an optomechanical
coupling $g/\omega_1 = 0.01$ has been used and  the cavity-mode
resonance frequency is $\omega_c =0.475\, \omega_1$. The lowest energy anticrossing
corresponds to the resonance condition for the DCE. The higher energy one is the signature of the mirror-mirror interaction mediated by the virtual DCE photons. At the minimum energy splitting $2 \lambda^{01}_{10} \simeq 1,85 \times 10^{-2} \omega_1$, the
resulting states are well approximated by $|\psi_{3,4} \rangle \simeq (1/\sqrt{2}) (|1,0,0 \rangle \pm |0,1,0 \rangle)$.
In panel (b) we use $g/\omega_1 = 0.1$. In this case the cavity-mode
resonance frequency is $\omega_c =0.8\, \omega_1$. Also in this case, the  anticrossing
is the signature of the mirror-mirror
interaction mediated by the virtual DCE photons.  At the minimum energy splitting $2 \lambda^{01}_{10} \simeq 2.56 \times 10^{-2} \omega_1$, the
resulting states are well approximated by $|\psi_{2,3} \rangle \simeq (1/\sqrt{2}) (|1,0,0 \rangle \pm |0,1,0 \rangle)$.

\subsection{System dynamics under a single-tone continuous-wave mechanical drive: additional results}
\label{Subsec:D}
We start investigating the system dynamics at  $T=0$, with the system starting from its ground state, and introducing the excitation of mirror 1 by a single-tone continuous-wave mechanical drive
${\cal F}_1(t) = {\cal A} \cos{(\omega_{\rm d} t)}$, with $\omega_d = \omega_1$.
\begin{figure}
	\centering
	\includegraphics[width = 6 cm]{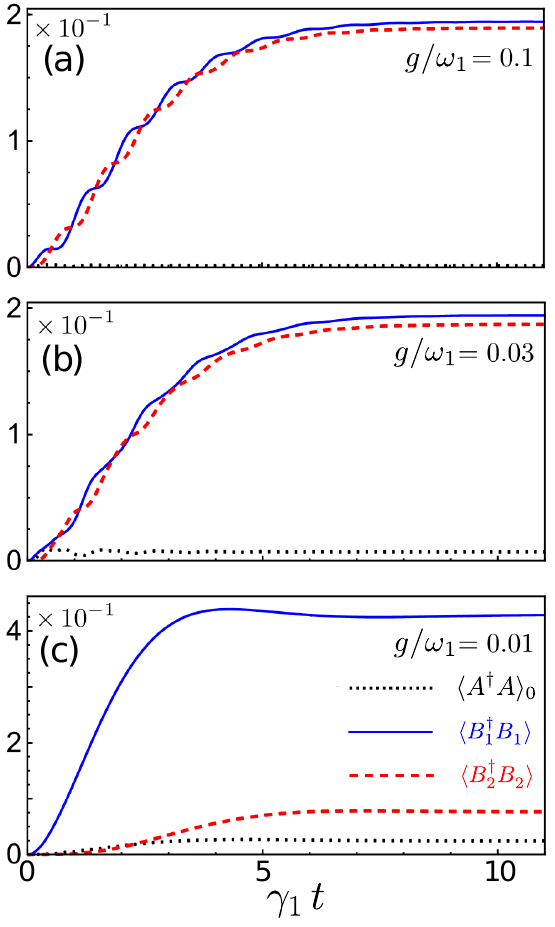}
	\caption{System dynamics under continuous-wave drive of  mirror 1 for different optomechanical coupling strengths. The blue solid and red dashed curves	describe   the mean phonon numbers $\langle \hat B_1^\dag \hat B_1^{} \rangle$ and $\langle \hat B_2^\dag \hat B_2^{} \rangle$, respectively, while the black dotted curve describes the mean intracavity photon number $\langle \hat A^\dag \hat A \rangle$. Parameters are given in the text.
		\label{figS:3}}
\end{figure}
Figure~\ref{figS:3} shows the time evolution of the mean phonon numbers of the two mirrors $\langle \hat B_i^\dag \hat B_i \rangle$ and of the  intracavity mean photon number $\langle \hat A^\dag \hat A \rangle$. Here $\hat A, \hat B_i$ are the {\em physical} photon and phonon operators (see main paper).
We assume a zero-temperature reservoir and use  $\gamma_1 = \gamma_2 = \gamma=  \omega_1 /260$ and $\kappa =  \gamma$ for the mechanical and photonic  loss rates. We consider a weak (${\cal A}/\gamma = 0.95$) resonant excitation of  mirror 1.
Panel (a) has been obtained using $g/\omega_1=0.1$ and $\omega_c / \omega_1 = 0.8$. Panel (b) using $g/\omega_1=0.03$ and $\omega_c / \omega_1 = 0.495$. Panel (c) using $g/\omega_1=0.01$ and $\omega_c / \omega_1 = 0.475$.
We set $\omega_2 = \omega_1$. The results shown in Fig.~\ref{figS:3} demonstrate that the excitation transfer mechanism via  virtual DCE photon pairs, proposed here, works properly. In steady state, mirror 2 reaches almost the same excitation intensity as the driven mirror 1 at normalized couplings $g=0.1$ and $g=0.03$. 
The photon population remains very low throughout the considered time window.
\begin{figure}[H]
	\centering
	\includegraphics[width = 0.9\linewidth]{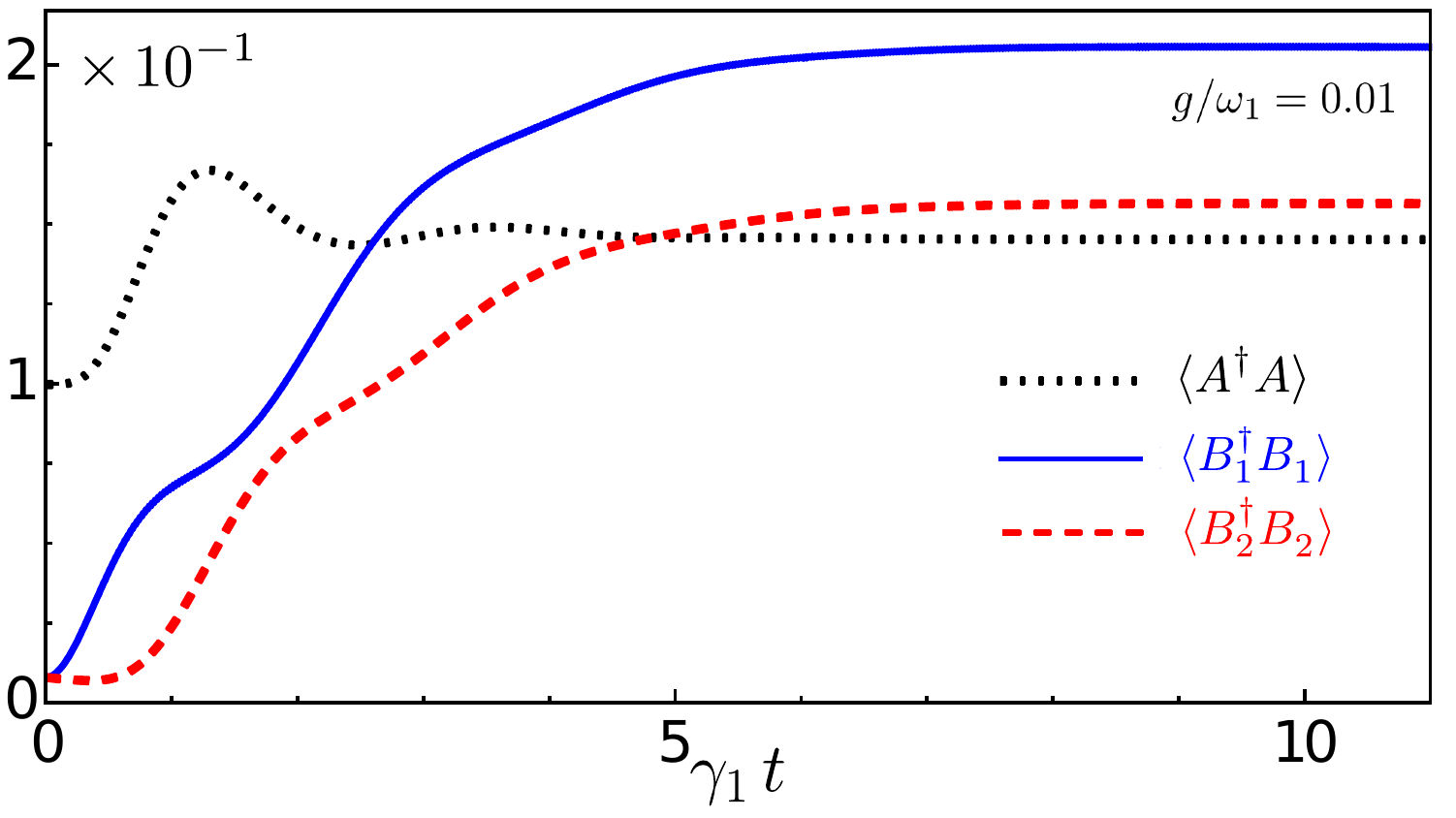}
	\caption{System dynamics for $\omega_{\rm c} = 0.5 \omega_{1}$ under continuous-wave drive of  mirror 1, normalized coupling  $g/ \omega_{1} = 0.01$ and T= $60$ mK. The blue solid and red dashed curves
		describe   the mean phonon numbers $\langle \hat B_1^\dag \hat B_1^{} \rangle$ and $\langle \hat B_2^\dag \hat B_2^{} \rangle$, respectively, while the black dotted curve describes the mean intracavity photon number $\langle \hat A^\dag \hat A \rangle$ arising due to the DCE.
		\label{figS:4}}
\end{figure}
In Fig.~\ref{figS:4}, in order to obtain the maximum excitation transfer between the two mirrors (despite the small coupling strength $g/ \omega_{1} = 0.01$), we investigate the system dynamics using $\omega_c=0.5\, \omega_1$. 
We also consider the system  initially in a thermal state  with a normalized thermal energy $k T / \omega_1 = 0.208$, corresponding to a temperature $T = 60$ mK for $\omega_1 / 2 \pi = 6$ GHz. During, its time evolution, the system interacts with  thermal baths with the same temperature $T$.  
The obtained results show that a good mechanical transfer is achieved.
However, in this case, a significant amount of real photon pairs are generated. This configuration can be used to probe the DCE effect in the presence of thermal photons.

\subsection{Mechanical excitation transfer: pulsed excitation}
\label{Subsec:E}

We now investigate the transfer of mechanical excitations  mediated by virtual photon pairs  
by exciting mirror 1 with a resonant Gaussian pulse:
$${\cal F}_1(t) = {\cal A}\, {\cal G}(t-t_0) \cos{(\omega_{\rm d}\, t)},$$
where $\omega_{\rm d} = \omega_1$, and ${\cal G}(t)$ is a normalized Gaussian function with standard deviation $\sigma = 1 / (10 \lambda^{01}_{10})$. 
\begin{figure}
	\centering
	\includegraphics[width = 8.5 cm]{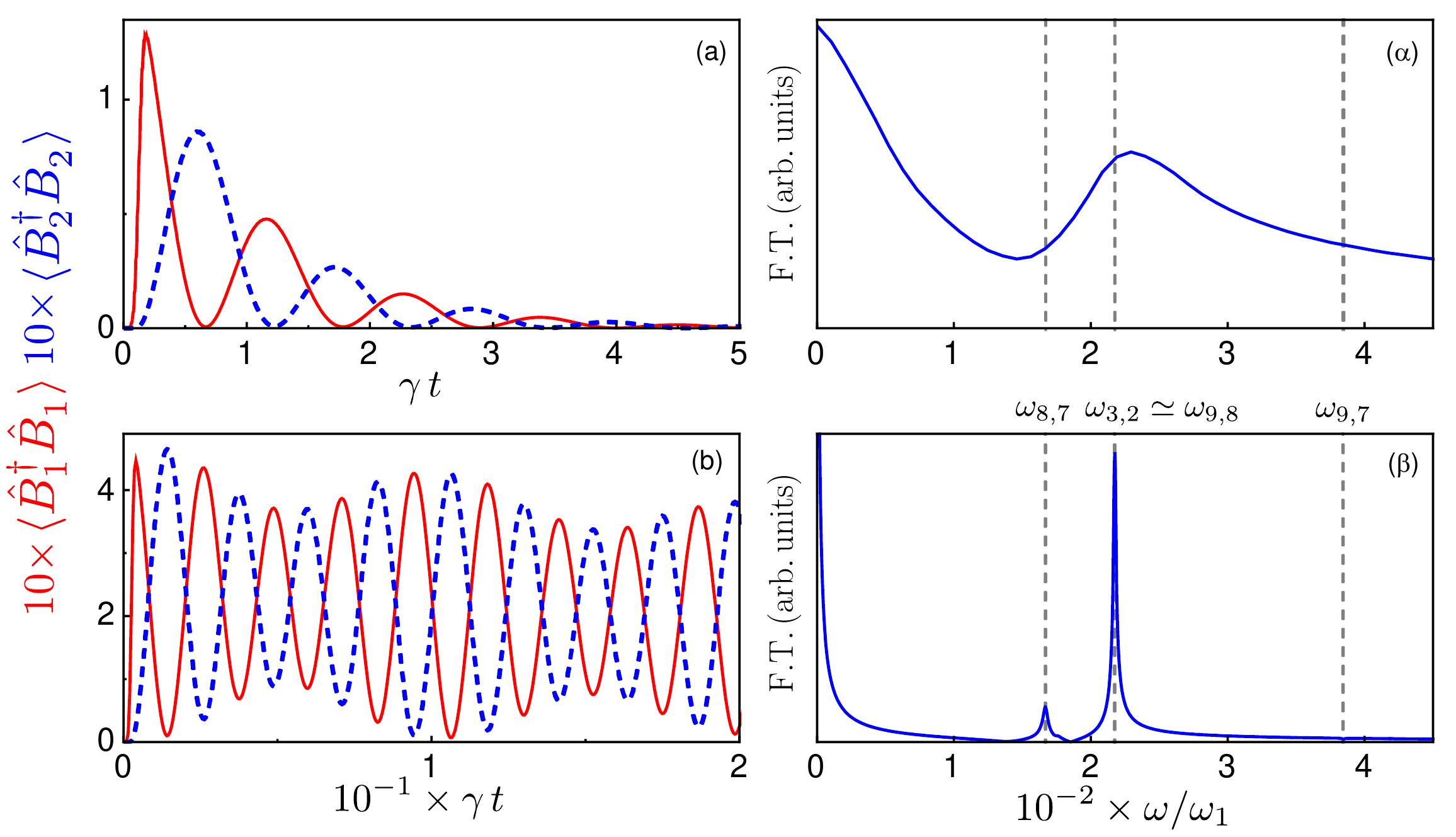}
	\caption{Time evolution of the mean phonon numbers of the two mirrors  after the  arrival of the pulse. We consider two different amplitudes which increase from top to bottom: ${\cal A} = 0.25\,\pi$ (a), $0.45 \pi$ (b). Specifically, panels (a-b) display the mean phonon numbers $\langle \hat B_i^\dag \hat B_i^{} \rangle$. Panels ($\alpha$-$\beta$) display the Fourier transform of the mean phonon number shown in the corresponding panel on the left. Other parameters are given in the text.
		\label{fig:3S}}
\end{figure}
We consider the case of the strong coupling regime, when the mirror-mirror coupling strength $\lambda_{10}^{01}$ is larger than the total decoherence rate $\gamma_1 + \gamma_2$.
We set the resonance frequency of mirror 2 to $\omega_2 \simeq \omega_1$ providing the minimum level splitting $2 \lambda_{10}^{01}$. The system starts in its ground state. Figure~\ref{fig:3S} displays the system dynamics after the pulse arrival and the Fourier transform of the mean phonon number of mirror 1 (no relevant changes occur for mirror 2), obtained for pulses with amplitudes increasing from top to bottom: ${\cal A} = 0.25 \pi, 0.45 \pi$.
Panels \ref{fig:3S}(a) and \ref{fig:3S}($\alpha$) have been obtained using the loss rates $\gamma = 3.5 \times 10^{-3} \omega_1$ and $\kappa= 0.5 \gamma$.  Figure \ref{fig:3S}(a) displays coherent and reversible sinusoidal oscillations (with peak amplitudes decaying exponentially), showing that the mechanical state  of the spatially separated mirrors is transferred from one to the other at a rate $\omega_{3,2} \equiv E_3 - E_2 = \lambda_{10}^{01}$, as confirmed by the peak in the Fourier transform in Fig.~\ref{fig:3S}($\alpha$).
We notice that the position and broadening  of the peak at $\omega_{3,2}$ in Fig.~\ref{fig:3S}($\alpha$) is influenced by the initial dynamics of $\langle \hat B^\dag_1 \hat B_1 \rangle$, which in turn is affected by the pulse shape (\figref{M2FT} displays the corresponding spectrum for mirror 2).
The higher peak at $\omega = 0$ originates from the exponential decay of the signal.
These results clearly show that, for the weaker excitation amplitude (${\cal A} = 0.25 \pi$), only the one-phonon states $|1,0,0 \rangle$ and $|0,1,0\rangle$ are excited significantly and contribute to the dynamics.

\begin{figure}[H]
	\centering
	\includegraphics[width = 8 cm]{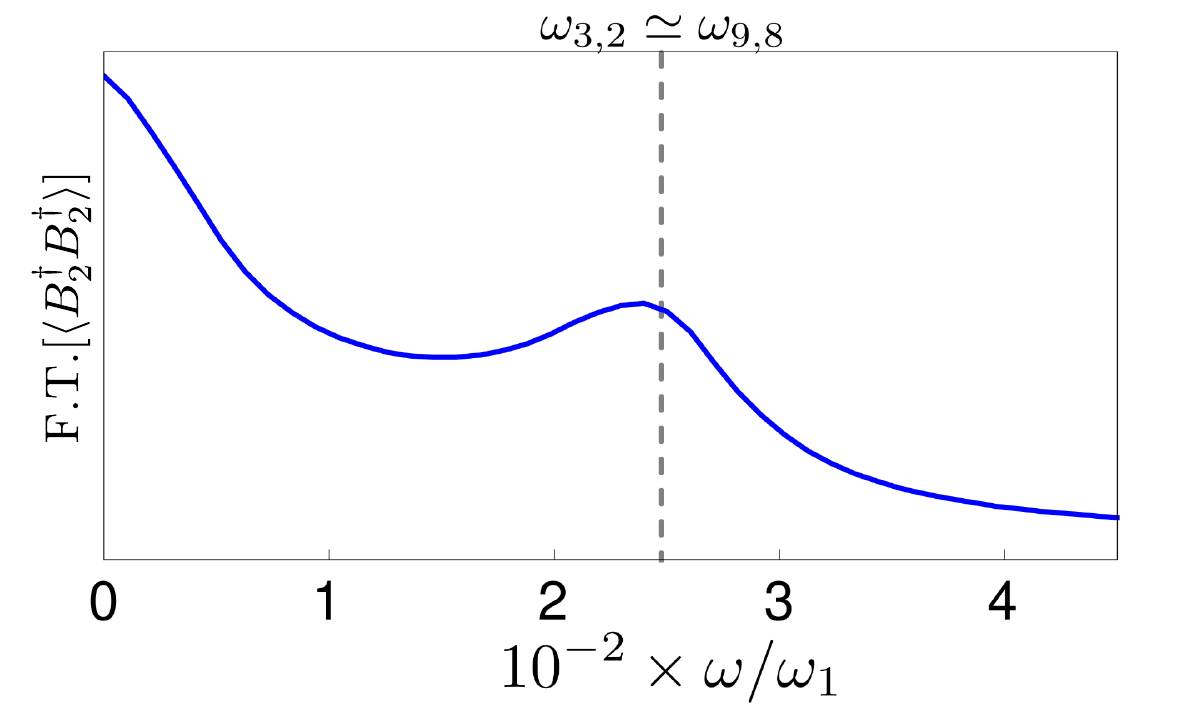} 
	\caption{
		Fourier transform of the mean phonon number of mirror 2  obtained for a pulse with amplitude ${\cal A} = 0.25 \pi$.}\label{M2FT}
\end{figure}

By increasing the pulse amplitude [Fig.~\ref{fig:3S}(b)], the mean phonon numbers grow significantly and the signals are no more sinusoidal, owing to the additional excitation of the states $|2,0,0 \rangle$, $|1,1,0 \rangle$, and $|0,2,0 \rangle$, whose DCE-induced coupling gives rise to the hybridized energy eigenstates $|\psi_7 \rangle$, $|\psi_8 \rangle$, and $|\psi_9 \rangle$. In order to better distinguish the nonsinusoidal behaviour, we used much lower loss rates:  $\gamma = 8 \times 10^{-5} \omega$ and $\kappa= 0.5 \gamma$. Figure~\ref{fig:3S}($\beta$) shows the appearence of an additional peak at $\omega= \omega_{8,7}$, confirming that higher-energy mechanical states get excited. We observe that the frequency splitting $\omega_{9,8}$ is very close to $\omega_{3,2}$, hence, it does not give rise to a new peak in Fig.~\ref{fig:3S}($\beta$). Moreover, the
frequency splitting at $\omega_{9,7}$ does not contribute significantly to the dynamics as confirmed by the spectrum in Fig.~\ref{fig:3S}($\beta$). An analytic calculation based on three coupled levels confirms that the used parameters give rise to a negligible contribution at $\omega_{9,7}$.

\begin{figure} [h!]
	\centering
	\includegraphics[width = 7 cm]{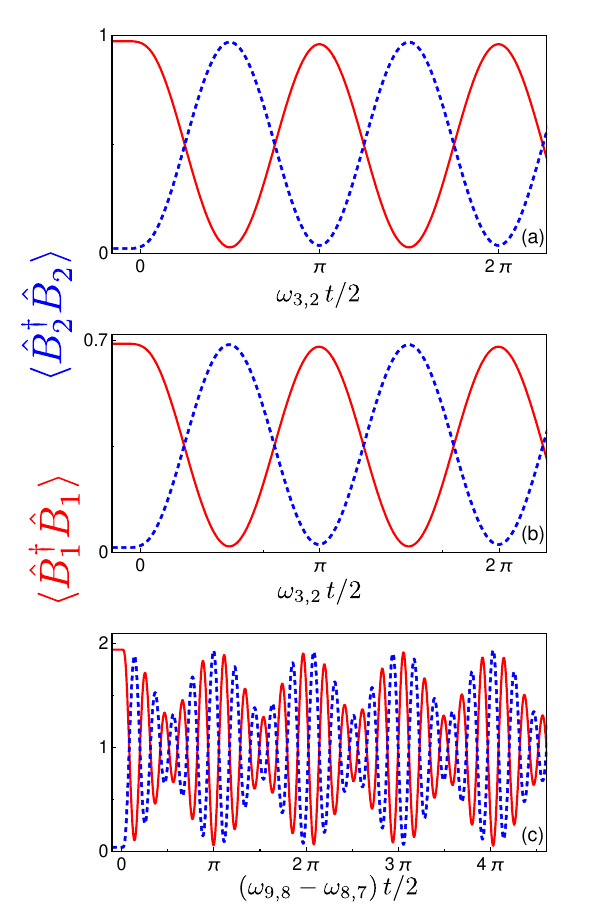}
	\caption{
		Time evolution of the mean phonon numbers of the two mirrors obtained preparing the system in an initial state 
		(a) $\ket{1,0,0}$, (b) $\frac{1}{\sqrt{2}}(\ket{1,0,0}+\ket{0,0,1})$, (c) $\ket{2,0,0}$.
		Mirror 2 is initially  set at a mechanical frequency $\omega^{\rm in}_2$ (details are given in the text). We note that  the dynamics display oscillations, (a) and (b),  due to 
		the avoided level crossing between the states $\ket{\psi_3}$ and $\ket{\psi_2}$ with frequency equal to  $\omega_{3,2}$; (c) due to the splittings between the states $\ket{\psi_9}$,  $\ket{\psi_8}$ and $\ket{\psi_7}$, whose transitions from higher to lower levels give rise to beats (the details are given in the text).
	}\label{nonadiabaticnodamping}
\end{figure}

\subsection{Mechanical excitation transfer: nonadiabatic effective switching of the interaction }
\label{Subsec:F}
\begin{figure}[h!]
	\centering
	\includegraphics[width = 7 cm]{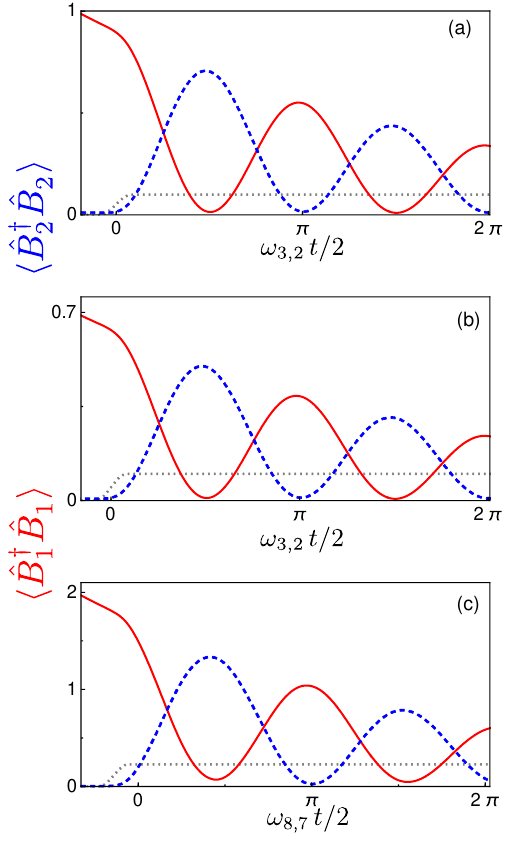}
	\caption{Time evolution of the mean phonon numbers of the two mirrors calculated after a non-adiabatic switching of the interaction, as explained in  \figref{nonadiabaticnodamping}, but in the presence of losses both in mirrors and cavity. The parameters are the same as in \figref{nonadiabaticnodamping}; in addition we have $\gamma=\gamma_1=\gamma_2=\omega_1/650$ and $\kappa=0.5\, \gamma$.
		The system is initially preparated in the states (a) $\ket{1,0,0}$, (b) $\frac{1}{\sqrt{2}}(\ket{1,0,0}+\ket{0,0,1})$, (c) $\ket{2,0,0}$. As we can observe, the oscillations are damped and disappear after a few periods. In (c) the losses do not allow for observations of  beats oscillations having a longer time period. The dotted gray lines show how the frequency of mirror 2 is tuned into resonance with mirror 1 (details are given in the text).
	}\label{nonadiabaticwithdamping}
\end{figure}

As pointed out in the last paragraph of the main paper, if it is possible to control the interaction time (as currently realized in superconducting artificial atoms), e.g., by rapidly changing  the resonance frequencies of the mechanical oscillators, the interaction scheme  proposed here would represent an attractive architecture for quantum information processing with optomechanical systems. Here we provide some examples of quantum state transfer.
In  \figref{nonadiabaticnodamping}, we show the phonon population dynamics obtained preparing the system in three different initial  states
(a) $\ket{1,0,0}$, (b) $\frac{1}{\sqrt{2}}(\ket{0,0,0}+\ket{1,0,0})$, (c) $\ket{2,0,0}$.
Mirror 2 is initially  set at a mechanical frequency $\omega^{\rm in}_2$. This value must be chosen sufficiently far from the value $\omega_2^{\rm min}\simeq \, 0.99\, \omega_1$ corresponding to the minimum splitting between states $\ket{1,0,0}$ and $\ket{0,1,0}$.
In particular, we have fixed   $\omega^{\rm in}_2=\omega_2^{\rm min}- \delta $ with $\delta=0.069\, \omega_1 $. This value is also sufficiently far from the region where the avoided three-level crossing between the states $\ket{\psi_i}$ with $i=7,8,9$ appears. Subsequently, a time-dependent perturbation $H_{\rm na}= f(t) \hat B_2^{\dag}\hat B_2$ [with $f(t)\approx \theta(t-t_0)$]   is introduced in order to modify the resonance frequency of mirror 2 ($\theta$ is the Heaviside step function). More specifically 
$
f(t)=\delta\left[\sin^2[\Omega(t-t_0)\theta(t-t_0)+\sin^2[\Omega(t-t_f)\theta(t-t_f)]\right]
$ is a smoothed step function, where $\delta$ fixes the change in  mechanical frequency of mirror 2, $t_0$ is the time when the frequency starts to change, $t_f= t_0+\pi/(2A)$, and $\Omega$ is the frequency setting the smoothness.

This enables a non-adiabatic transition from the frequency region with $\omega_2=\omega^{\rm in}_2$, where the states $\ket{2,0,0}$, $\ket{1,0,0}$ and $\ket{0,1,0}$ are eigenstates of the system, to the frequency region $\omega_2=\omega_2^{\rm min}$ where the former states are no longer eigenstates of the system. As a consequence,  the dynamics of the phonon populations of the two mirrors display  {\em quantum Rabi-like} oscillations [see \figref{nonadiabaticnodamping}(a) and (b)] due to 
the avoided level crossing between the states $\ket{\psi_3}$ and $\ket{\psi_2}$  (the eigenstates of the systems  are, in this frequency region,  the symmetric and antisymmetric superpositions of $\ket{1,0,0}$ and $\ket{0,1,0}$; see Fig.~1b in the main paper). In  \figref{nonadiabaticnodamping}(c), the avoided level crossing between the states $\ket{\psi_9}$,  $\ket{\psi_8}$, and $\ket{\psi_7}$ gives rise to transitions from higher to lower levels. As a consequence,  we observe beats between the two transition frequencies $\omega_{8,9}$ and $\omega_{8,7}$ (with the chosen parameters the other frequency transition $\omega_{9,7}$ does not contribute to the beats). 
Finally, in  \figref{nonadiabaticwithdamping}, we show the time evolution of the mean phonon numbers for the same cases discussed above, but in the presence of losses both in mirrors and cavity. We observe the damping of the population dynamics as expected in presence of losses.

\subsection{Experimental platform for the observation of the proposed effect} \label{Subsec:G}

A platform to experimentally demonstrate these results is circuit optomechanics using ultra-high-frequency ($\omega_{\rm 1}$ at 4-6 GHz) dilatational resonators \cite{OConnell2010}. These mechanical oscillators have a resonance frequency  $f_m = v/2d$, where $v$ is the average  speed of sound and $d$ is the resonator thickness.
Their resonant quantum interaction with a superconducting phase qubit, described by the quantum Rabi  (or also the Jaynes-Cummings) Hamiltonian, has been experimentally demonstrated \cite{OConnell2010, Rouxinol2016}.
In the present case, we want to estimate the radiation-pressure interaction strength between the high-frequency mechanical resonator and an electromagnetic resonator. In order to estimate the achievable coupling strength, we begin by analyzing the coupling between a mechanical resonator and a flux qubit, experimentally realized in Ref.~\cite{OConnell2010}. Then we use the experimentally achieved qubit-oscillator coupling strength to derive an accurate estimate of the presently achievable radiation-pressure coupling strength between this mechanical resonator and an electromagnetic resonator. 
Note that the mechanical oscillator considered in Ref.~\cite{OConnell2010} has a quality factor equal to that used in our calculations: $Q_m = 260$. Moreover, it has been shown that lowering $f_m$ can strongly increase the quality factor \cite{Cleland2004}.

The mechanical resonator is coupled to a superconducting artificial atom through a capacitor \cite{OConnell2010}. An elastic strain in the vibrational resonator produces, through the piezoelectric effect, a charge on the capacitor enclosing it, which results in a charge $Q$ on the coupling capacitor giving a current $\dot {Q}$. 
The coupling energy is $\hat V' = (\hbar /2e)\,  \hat \varphi\,  \dot {\hat Q}$, where $\hat \varphi$ is the phase-difference operator of the Josephson junction. Considering only the two lowest energy levels (qubit) of the artificial atom, the phase operator can be expanded as $\hat \phi = (2 E'_C / E'_J)^{1/4}\, \hat \sigma_x$, resulting in the Rabi-like interaction Hamiltonian
\be\label{Vqm}
\hat V'_{qm} = \hbar \,  (2 E'_C / E'_J)^{1/4}\, \hat \sigma_x\,  (\dot {\hat Q}/2e)\, ,
\ee
where  $E'_C$ and $E'_J$ are the charging energy and the Josephson energy, respectively, of the phase qubit (with $E'_C \ll E'_J$), and $\dot {\hat Q}$ is proportional to the vibrational strain velocity $\dot {\hat x} = i \omega_1 X_{\rm zpf}\,  (\hat b_1^\dag - \hat b_1)$ ($X_{\rm zpf}$ is the zero-point fluctuation amplitude of the mechanical coordinate). Finally, this interaction Hamiltonian can also be expressed in the standard Rabi interaction form:
\be
\hat V'_{qm} = -i g'_m (\hat b - \hat b^\dag) \hat \sigma_x\, , 
\ee
where $g'_m$ is the resulting coupling strength and $\hat b$ and $\hat b^\dag$ are, respectively, the annihilation and creation operators for a generic mechanical oscillator.

For the observation of the effects described in this paper, optomechanical systems displaying a radiation-pressure interaction Hamiltonian are required.
Moreover a strong optomechanical coupling (at least $g/\omega_1 \sim 0.01$) is needed. This kind of interaction with a reasonable coupling strength can be obtained by considering a tripartite system consisting of an electromagnetic resonator, an ultra-high-frequency mechanical resonator, and a superconducting  charge qubit mediating the interaction between the former two parts \cite{Heikkila2014,Pirkkalainen2015}. It has been shown that the presence of the qubit can strongly enhance the optomechanical coupling.

Without presenting a detailed circuit-optomechanical setup, which goes beyond the scope of the present work, we can provide an estimate of the resulting coupling strength which can be achieved within state-of-the-art technology.
Specifically, considering one generic mechanical oscillator, coupled through a capacitor to a charge qubit, the  qubit-mechanical oscillator interaction Hamiltonian can be  written as $\hat V_{qm} = 8 E_C\, \hat n\, (\hat Q/2e)$, where 
$\hat n$ is the number operator for the Cooper pairs transferred across the junction. In the full charge qubit limit, $E_J \ll E_C$, the bare qubit transition energy is $\omega_q \approx 4 E_C$, and the mechanical coupling is longitudinal, i.e., in the two-state representation $\hat n \to \hat \sigma_z/2$. The resulting
interaction Hamiltonian is
\be\label{Vqm1}
\hat V_{qm} =  \hbar \omega_q\, \hat \sigma_z  (\hat Q/2e)\, ,
\ee
which can also be expressed as
\be\label{Vqm3}
\hat V_{qm} = g_m (\hat b+ \hat b^\dag) \hat \sigma_z\, .
\ee
Assuming that the same mechanical oscillator is coupled through the same capacitor to the two different kinds of superconducting qubits, it is possible to compare the two qubit-mechanical oscillator coupling strengths. From Eqs~(\ref{Vqm}) and (\ref{Vqm1}), disregarding the phase difference, we obtain
\be 
\frac{g_m}{g'_m} =   \left(\frac{ E'_J}{2E'_C}\right)^{\frac{1}{4}}
\frac{ \omega_q}{\omega_1}\, .
\ee
Below we will consider the case $2 \omega_q \sim \omega_1$. Assuming the energies $E'_J$ and $E'_C$ for a typical phase qubit (see, e.g. Ref.~\cite{Cleland2004}), we obtain $g_m /g'_m  \gtrsim 12$.

Now, following Refs.~\cite{Heikkila2014} and \cite{Pirkkalainen2015}, we consider the additional interaction of the charge qubit with an electromagnetic resonator, described by the Hamiltonian
\be
\hat V_{\rm qc} = g_c (\hat a+ \hat a^\dag) \hat \sigma_x\, ,
\ee
where $\hat a$ is the destruction operator of the cavity mode.
In the dispersive regime,  the qubit-cavity interaction can be well approximated by \cite{Zueco2009}
\be\label{V4}
\hat V_{\rm qc} = (g^2_c / 2 \Delta) \hat \sigma_z (\hat a +\hat a^\dagger)^2\, ,
\ee
where $\Delta = \omega_q -\omega_c$.  Corrections of the qubit energy not depending on photon operators have been disregarded. Equation~(\ref{Vqm3}) shows that the coupling of the charge qubit with the mechanical oscillator induces a qubit energy shift depending on the mechanical displacement, so that $\omega_q \to \omega_q + 2 g_m (\hat b + \hat b^\dag)$. Replacing $\Delta$ with $\Delta(\hat x) = \omega_q + 2 g_m (\hat b + \hat b^\dag) - \omega_c$ in Eq.~(\ref{V4}), assuming small displacements, and considering the qubit in its ground state, from Eq.~(\ref{V4}) we obtain the following optomechanical interaction,
\be 
\hat H_I = \frac{g}{2} (\hat a +\hat a^\dagger)^2 (\hat b + \hat b^\dag)\, ,
\ee
with
\be g = \frac{2 g_m g_c^2}{\Delta^2}\, .
\ee

Using $g_m = 0.02\, \omega_1$, corresponding to the  value  of the electromechanical system 
employed for the demonstration of single-phonon control of a mechanical resonator \cite{OConnell2010}, assuming $g_m/g'_m =12$, and considering a  detuning $\Delta = 5 g_c$, we obtain $g \simeq 0.02 \omega_1$.
The achievable value could be even  higher, noting that  the electromechanical system used in Ref.~\cite{OConnell2010} was designed to limit $g_m$ in order to optimize the transfer process \cite{Cleland2004}.

Beyond the direct observation of the energy transfer between the mechanical oscillators (see Fig.~1 in the main text), the effective coherent coupling between the two mirrors can  also be demonstrated by looking at the system response (e.g., $\langle \hat B^\dag_1 \hat B_1 \rangle$ ) under continuous-wave weak excitation as a function of the excitation frequency. For $\omega_1 = \omega_2$, if $\lambda > \gamma$, two  peaks should be observed, corresponding, e.g.,  to the avoided level crossing at higher energy in Fig.~\ref{fig:2}(b) in the main text or to that in Fig.~\ref{Sg01001}(a). In order to confirm that the two observed peaks originate from virtual DCE photons, it would be useful to perform measurements changing the optomechanical coupling. This coupling can be tuned by modifying the gate charge of the qubit mediating the interaction \cite{Pirkkalainen2015}. If the energy splitting originates from virtual DCE photons, as predicted by Eq.~(\ref{lambda01}) in the main text, it should grow quadratically with the optomechanical coupling $g$ (see Fig.~\ref{Fig2}). The anticrossing behaviour could also be probed, changing $d$ of one of the two dilatational resonators  and detecting, e.g., $\langle \hat B^\dag_1 \hat B_1 \rangle$ at steady state as a function of the thickness $d$ (note that $\omega_2 = v/d$). Two peaks with a splitting determined by the thickness, following the avoided level crossing   should be observed (see, e.g., Fig.~\ref{Sg01001}). The detection of the mechanical excitations can be performed following the procedures used in Refs.~\cite{Cleland2004,OConnell2014}.
\newline
\acknowledgments{FN is partially supported by the MURI Center for Dynamic Magneto-Optics via the AFOSR Award No.~FA9550-14-1-0040, the Army Research Office (ARO) under grant number W911NF-18-1-0358, the AOARD grant No.~ FA2386-18-1-4045, the CREST Grant No.~JPMJCR1676, the IMPACT program of JST, the RIKEN-AIST Challenge Research Fund, the JSPS-RFBR grant No.~17-52-50023, and the Sir John Templeton Foundation.}	
\bibliography{DCE}

\begin{thebibliography}{56}%
\makeatletter
\providecommand \@ifxundefined [1]{%
 \@ifx{#1\undefined}
}%
\providecommand \@ifnum [1]{%
 \ifnum #1\expandafter \@firstoftwo
 \else \expandafter \@secondoftwo
 \fi
}%
\providecommand \@ifx [1]{%
 \ifx #1\expandafter \@firstoftwo
 \else \expandafter \@secondoftwo
 \fi
}%
\providecommand \natexlab [1]{#1}%
\providecommand \enquote  [1]{``#1''}%
\providecommand \bibnamefont  [1]{#1}%
\providecommand \bibfnamefont [1]{#1}%
\providecommand \citenamefont [1]{#1}%
\providecommand \href@noop [0]{\@secondoftwo}%
\providecommand \href [0]{\begingroup \@sanitize@url \@href}%
\providecommand \@href[1]{\@@startlink{#1}\@@href}%
\providecommand \@@href[1]{\endgroup#1\@@endlink}%
\providecommand \@sanitize@url [0]{\catcode `\\12\catcode `\$12\catcode
  `\&12\catcode `\#12\catcode `\^12\catcode `\_12\catcode `\%12\relax}%
\providecommand \@@startlink[1]{}%
\providecommand \@@endlink[0]{}%
\providecommand \url  [0]{\begingroup\@sanitize@url \@url }%
\providecommand \@url [1]{\endgroup\@href {#1}{\urlprefix }}%
\providecommand \urlprefix  [0]{URL }%
\providecommand \Eprint [0]{\href }%
\providecommand \doibase [0]{http://dx.doi.org/}%
\providecommand \selectlanguage [0]{\@gobble}%
\providecommand \bibinfo  [0]{\@secondoftwo}%
\providecommand \bibfield  [0]{\@secondoftwo}%
\providecommand \translation [1]{[#1]}%
\providecommand \BibitemOpen [0]{}%
\providecommand \bibitemStop [0]{}%
\providecommand \bibitemNoStop [0]{.\EOS\space}%
\providecommand \EOS [0]{\spacefactor3000\relax}%
\providecommand \BibitemShut  [1]{\csname bibitem#1\endcsname}%
\let\auto@bib@innerbib\@empty
\bibitem [{\citenamefont {Majer}\ \emph {et~al.}(2007)\citenamefont {Majer},
  \citenamefont {Chow}, \citenamefont {Gambetta}, \citenamefont {Koch},
  \citenamefont {Johnson}, \citenamefont {Schreier}, \citenamefont {Frunzio},
  \citenamefont {Schuster}, \citenamefont {Houck}, \citenamefont {Wallraff},
  \citenamefont {Blais}, \citenamefont {Devoret}, \citenamefont {Girvin},\ and\
  \citenamefont {Schoelkopf}}]{Majer2007}%
  \BibitemOpen
  \bibfield  {author} {\bibinfo {author} {\bibfnamefont {J.}~\bibnamefont
  {Majer}}, \bibinfo {author} {\bibfnamefont {J.~M.}\ \bibnamefont {Chow}},
  \bibinfo {author} {\bibfnamefont {J.~M.}\ \bibnamefont {Gambetta}}, \bibinfo
  {author} {\bibfnamefont {J.}~\bibnamefont {Koch}}, \bibinfo {author}
  {\bibfnamefont {B.~R.}\ \bibnamefont {Johnson}}, \bibinfo {author}
  {\bibfnamefont {J.~A.}\ \bibnamefont {Schreier}}, \bibinfo {author}
  {\bibfnamefont {L.}~\bibnamefont {Frunzio}}, \bibinfo {author} {\bibfnamefont
  {D.~I.}\ \bibnamefont {Schuster}}, \bibinfo {author} {\bibfnamefont {A.~A.}\
  \bibnamefont {Houck}}, \bibinfo {author} {\bibfnamefont {A.}~\bibnamefont
  {Wallraff}}, \bibinfo {author} {\bibfnamefont {A.}~\bibnamefont {Blais}},
  \bibinfo {author} {\bibfnamefont {M.~H.}\ \bibnamefont {Devoret}}, \bibinfo
  {author} {\bibfnamefont {S.~M.}\ \bibnamefont {Girvin}}, \ and\ \bibinfo
  {author} {\bibfnamefont {R.~J.}\ \bibnamefont {Schoelkopf}},\ }\bibfield
  {title} {\enquote {\bibinfo {title} {Coupling superconducting qubits via a
  cavity bus},}\ }\href {\doibase 10.1038/nature06184} {\bibfield  {journal}
  {\bibinfo  {journal} {Nature}\ }\textbf {\bibinfo {volume} {449}},\ \bibinfo
  {pages} {443} (\bibinfo {year} {2007})}\BibitemShut {NoStop}%
\bibitem [{\citenamefont {S\o{}rensen}\ and\ \citenamefont
  {M\o{}lmer}(1999)}]{Sorensen1999}%
  \BibitemOpen
  \bibfield  {author} {\bibinfo {author} {\bibfnamefont {A.}~\bibnamefont
  {S\o{}rensen}}\ and\ \bibinfo {author} {\bibfnamefont {K.}~\bibnamefont
  {M\o{}lmer}},\ }\bibfield  {title} {\enquote {\bibinfo {title} {Quantum
  computation with ions in thermal motion},}\ }\href {\doibase
  10.1103/PhysRevLett.82.1971} {\bibfield  {journal} {\bibinfo  {journal}
  {Phys. Rev. Lett.}\ }\textbf {\bibinfo {volume} {82}},\ \bibinfo {pages}
  {1971} (\bibinfo {year} {1999})}\BibitemShut {NoStop}%
\bibitem [{\citenamefont {Imamo\={g}lu}\ \emph {et~al.}(1999)\citenamefont
  {Imamo\={g}lu}, \citenamefont {Awschalom}, \citenamefont {Burkard},
  \citenamefont {DiVincenzo}, \citenamefont {Loss}, \citenamefont {Sherwin},\
  and\ \citenamefont {Small}}]{Imamoglu1999}%
  \BibitemOpen
  \bibfield  {author} {\bibinfo {author} {\bibfnamefont {A.}~\bibnamefont
  {Imamo\={g}lu}}, \bibinfo {author} {\bibfnamefont {D.~D.}\ \bibnamefont
  {Awschalom}}, \bibinfo {author} {\bibfnamefont {G.}~\bibnamefont {Burkard}},
  \bibinfo {author} {\bibfnamefont {D.~P.}\ \bibnamefont {DiVincenzo}},
  \bibinfo {author} {\bibfnamefont {D.}~\bibnamefont {Loss}}, \bibinfo {author}
  {\bibfnamefont {M.}~\bibnamefont {Sherwin}}, \ and\ \bibinfo {author}
  {\bibfnamefont {A.}~\bibnamefont {Small}},\ }\bibfield  {title} {\enquote
  {\bibinfo {title} {Quantum information processing using quantum dot spins and
  cavity {QED}},}\ }\href {\doibase 10.1103/PhysRevLett.83.4204} {\bibfield
  {journal} {\bibinfo  {journal} {Phys. Rev. Lett.}\ }\textbf {\bibinfo
  {volume} {83}},\ \bibinfo {pages} {4204} (\bibinfo {year}
  {1999})}\BibitemShut {NoStop}%
\bibitem [{\citenamefont {Zheng}\ and\ \citenamefont {Guo}(2000)}]{Zheng2000}%
  \BibitemOpen
  \bibfield  {author} {\bibinfo {author} {\bibfnamefont {S.-B.}\ \bibnamefont
  {Zheng}}\ and\ \bibinfo {author} {\bibfnamefont {G.-C.}\ \bibnamefont
  {Guo}},\ }\bibfield  {title} {\enquote {\bibinfo {title} {Efficient scheme
  for two-atom entanglement and quantum information processing in cavity
  {QED}},}\ }\href {\doibase 10.1103/PhysRevLett.85.2392} {\bibfield  {journal}
  {\bibinfo  {journal} {Phys. Rev. Lett.}\ }\textbf {\bibinfo {volume} {85}},\
  \bibinfo {pages} {2392} (\bibinfo {year} {2000})}\BibitemShut {NoStop}%
\bibitem [{\citenamefont {Osnaghi}\ \emph {et~al.}(2001)\citenamefont
  {Osnaghi}, \citenamefont {Bertet}, \citenamefont {Auffeves}, \citenamefont
  {Maioli}, \citenamefont {Brune}, \citenamefont {Raimond},\ and\ \citenamefont
  {Haroche}}]{osnaghi2001}%
  \BibitemOpen
  \bibfield  {author} {\bibinfo {author} {\bibfnamefont {S.}~\bibnamefont
  {Osnaghi}}, \bibinfo {author} {\bibfnamefont {P.}~\bibnamefont {Bertet}},
  \bibinfo {author} {\bibfnamefont {A.}~\bibnamefont {Auffeves}}, \bibinfo
  {author} {\bibfnamefont {P.}~\bibnamefont {Maioli}}, \bibinfo {author}
  {\bibfnamefont {M.}~\bibnamefont {Brune}}, \bibinfo {author} {\bibfnamefont
  {J.-M.}\ \bibnamefont {Raimond}}, \ and\ \bibinfo {author} {\bibfnamefont
  {S.}~\bibnamefont {Haroche}},\ }\bibfield  {title} {\enquote {\bibinfo
  {title} {Coherent control of an atomic collision in a cavity},}\ }\href
  {\doibase 10.1103/physrevlett.87.037902} {\bibfield  {journal} {\bibinfo
  {journal} {Phys. Rev. Lett.}\ }\textbf {\bibinfo {volume} {87}},\ \bibinfo
  {pages} {037902} (\bibinfo {year} {2001})}\BibitemShut {NoStop}%
\bibitem [{\citenamefont {Filipp}\ \emph {et~al.}(2011)\citenamefont {Filipp},
  \citenamefont {G{\"o}ppl}, \citenamefont {Fink}, \citenamefont {Baur},
  \citenamefont {Bianchetti}, \citenamefont {Steffen},\ and\ \citenamefont
  {Wallraff}}]{filipp2011}%
  \BibitemOpen
  \bibfield  {author} {\bibinfo {author} {\bibfnamefont {S.}~\bibnamefont
  {Filipp}}, \bibinfo {author} {\bibfnamefont {M.}~\bibnamefont {G{\"o}ppl}},
  \bibinfo {author} {\bibfnamefont {J.~M.}\ \bibnamefont {Fink}}, \bibinfo
  {author} {\bibfnamefont {M.}~\bibnamefont {Baur}}, \bibinfo {author}
  {\bibfnamefont {R.}~\bibnamefont {Bianchetti}}, \bibinfo {author}
  {\bibfnamefont {L.}~\bibnamefont {Steffen}}, \ and\ \bibinfo {author}
  {\bibfnamefont {A.}~\bibnamefont {Wallraff}},\ }\bibfield  {title} {\enquote
  {\bibinfo {title} {Multimode mediated qubit-qubit coupling and dark-state
  symmetries in circuit quantum electrodynamics},}\ }\href {\doibase
  10.1103/physreva.83.063827} {\bibfield  {journal} {\bibinfo  {journal}
  {Physical Review A}\ }\textbf {\bibinfo {volume} {83}},\ \bibinfo {pages}
  {063827} (\bibinfo {year} {2011})}\BibitemShut {NoStop}%
\bibitem [{\citenamefont {DiCarlo}\ \emph {et~al.}(2009)\citenamefont
  {DiCarlo}, \citenamefont {Chow}, \citenamefont {Gambetta}, \citenamefont
  {Bishop}, \citenamefont {Johnson}, \citenamefont {Schuster}, \citenamefont
  {Majer}, \citenamefont {Blais}, \citenamefont {Frunzio}, \citenamefont
  {Girvin},\ and\ \citenamefont {Schoelkopf}}]{DiCarlo2009}%
  \BibitemOpen
  \bibfield  {author} {\bibinfo {author} {\bibfnamefont {L.}~\bibnamefont
  {DiCarlo}}, \bibinfo {author} {\bibfnamefont {J.~M.}\ \bibnamefont {Chow}},
  \bibinfo {author} {\bibfnamefont {J.~M.}\ \bibnamefont {Gambetta}}, \bibinfo
  {author} {\bibfnamefont {L.~S.}\ \bibnamefont {Bishop}}, \bibinfo {author}
  {\bibfnamefont {B.~R.}\ \bibnamefont {Johnson}}, \bibinfo {author}
  {\bibfnamefont {D.~I.}\ \bibnamefont {Schuster}}, \bibinfo {author}
  {\bibfnamefont {J.}~\bibnamefont {Majer}}, \bibinfo {author} {\bibfnamefont
  {A.}~\bibnamefont {Blais}}, \bibinfo {author} {\bibfnamefont
  {L.}~\bibnamefont {Frunzio}}, \bibinfo {author} {\bibfnamefont {S.~M.}\
  \bibnamefont {Girvin}}, \ and\ \bibinfo {author} {\bibfnamefont {R.~J.}\
  \bibnamefont {Schoelkopf}},\ }\bibfield  {title} {\enquote {\bibinfo {title}
  {Demonstration of two-qubit algorithms with a superconducting quantum
  processor},}\ }\href
  {https://www.nature.com/nature/journal/v460/n7252/full/nature08121.html}
  {\bibfield  {journal} {\bibinfo  {journal} {Nature}\ }\textbf {\bibinfo
  {volume} {460}},\ \bibinfo {pages} {240} (\bibinfo {year}
  {2009})}\BibitemShut {NoStop}%
\bibitem [{\citenamefont {Neeley}\ \emph {et~al.}(2010)\citenamefont {Neeley},
  \citenamefont {Bialczak}, \citenamefont {Lenander}, \citenamefont {Lucero},
  \citenamefont {Mariantoni}, \citenamefont {O'Connell}, \citenamefont {Sank},
  \citenamefont {Wang}, \citenamefont {Weides}, \citenamefont {Wenner},
  \citenamefont {Yin}, \citenamefont {Yamamoto}, \citenamefont {Cleland},\ and\
  \citenamefont {Martinis}}]{Neeley2010}%
  \BibitemOpen
  \bibfield  {author} {\bibinfo {author} {\bibfnamefont {M.}~\bibnamefont
  {Neeley}}, \bibinfo {author} {\bibfnamefont {R.~C.}\ \bibnamefont
  {Bialczak}}, \bibinfo {author} {\bibfnamefont {M.}~\bibnamefont {Lenander}},
  \bibinfo {author} {\bibfnamefont {E.}~\bibnamefont {Lucero}}, \bibinfo
  {author} {\bibfnamefont {M.}~\bibnamefont {Mariantoni}}, \bibinfo {author}
  {\bibfnamefont {A.~D.}\ \bibnamefont {O'Connell}}, \bibinfo {author}
  {\bibfnamefont {D.}~\bibnamefont {Sank}}, \bibinfo {author} {\bibfnamefont
  {H.}~\bibnamefont {Wang}}, \bibinfo {author} {\bibfnamefont {M.}~\bibnamefont
  {Weides}}, \bibinfo {author} {\bibfnamefont {J.}~\bibnamefont {Wenner}},
  \bibinfo {author} {\bibfnamefont {Y.}~\bibnamefont {Yin}}, \bibinfo {author}
  {\bibfnamefont {T.}~\bibnamefont {Yamamoto}}, \bibinfo {author}
  {\bibfnamefont {A.~N.}\ \bibnamefont {Cleland}}, \ and\ \bibinfo {author}
  {\bibfnamefont {J.~M.}\ \bibnamefont {Martinis}},\ }\bibfield  {title}
  {\enquote {\bibinfo {title} {Generation of three-qubit entangled states using
  superconducting phase qubits},}\ }\href
  {https://www.nature.com/nature/journal/v467/n7315/full/nature09418.html}
  {\bibfield  {journal} {\bibinfo  {journal} {Nature}\ }\textbf {\bibinfo
  {volume} {467}},\ \bibinfo {pages} {570} (\bibinfo {year}
  {2010})}\BibitemShut {NoStop}%
\bibitem [{\citenamefont {Stassi}\ \emph {et~al.}(2017)\citenamefont {Stassi},
  \citenamefont {Macr{\`\i}}, \citenamefont {Kockum}, \citenamefont
  {Di~Stefano}, \citenamefont {Miranowicz}, \citenamefont {Savasta},\ and\
  \citenamefont {Nori}}]{Stassi2017}%
  \BibitemOpen
  \bibfield  {author} {\bibinfo {author} {\bibfnamefont {R.}~\bibnamefont
  {Stassi}}, \bibinfo {author} {\bibfnamefont {V.}~\bibnamefont {Macr{\`\i}}},
  \bibinfo {author} {\bibfnamefont {A.~F.}\ \bibnamefont {Kockum}}, \bibinfo
  {author} {\bibfnamefont {O.}~\bibnamefont {Di~Stefano}}, \bibinfo {author}
  {\bibfnamefont {A.}~\bibnamefont {Miranowicz}}, \bibinfo {author}
  {\bibfnamefont {S.}~\bibnamefont {Savasta}}, \ and\ \bibinfo {author}
  {\bibfnamefont {F.}~\bibnamefont {Nori}},\ }\bibfield  {title} {\enquote
  {\bibinfo {title} {Quantum nonlinear optics without photons},}\ }\href
  {\doibase 10.1103/physreva.96.023818} {\bibfield  {journal} {\bibinfo
  {journal} {Phys. Rev. A}\ }\textbf {\bibinfo {volume} {96}},\ \bibinfo
  {pages} {023818} (\bibinfo {year} {2017})}\BibitemShut {NoStop}%
\bibitem [{\citenamefont {Zhao}\ \emph {et~al.}(2017)\citenamefont {Zhao},
  \citenamefont {Tan}, \citenamefont {Yu}, \citenamefont {Zhu},\ and\
  \citenamefont {Yu}}]{Zhao2017}%
  \BibitemOpen
  \bibfield  {author} {\bibinfo {author} {\bibfnamefont {P.}~\bibnamefont
  {Zhao}}, \bibinfo {author} {\bibfnamefont {X.}~\bibnamefont {Tan}}, \bibinfo
  {author} {\bibfnamefont {H.}~\bibnamefont {Yu}}, \bibinfo {author}
  {\bibfnamefont {S.-L.}\ \bibnamefont {Zhu}}, \ and\ \bibinfo {author}
  {\bibfnamefont {Y.}~\bibnamefont {Yu}},\ }\bibfield  {title} {\enquote
  {\bibinfo {title} {Circuit {QED} with qutrits: Coupling three or more atoms
  via virtual-photon exchange},}\ }\href {\doibase 10.1103/PhysRevA.96.043833}
  {\bibfield  {journal} {\bibinfo  {journal} {Phys. Rev. A}\ }\textbf {\bibinfo
  {volume} {96}},\ \bibinfo {pages} {043833} (\bibinfo {year}
  {2017})}\BibitemShut {NoStop}%
\bibitem [{\citenamefont {Kockum}\ \emph
  {et~al.}(2017{\natexlab{a}})\citenamefont {Kockum}, \citenamefont
  {Miranowicz}, \citenamefont {Macr\`{\i}}, \citenamefont {Savasta},\ and\
  \citenamefont {Nori}}]{Kockum2017}%
  \BibitemOpen
  \bibfield  {author} {\bibinfo {author} {\bibfnamefont {A.~F.}\ \bibnamefont
  {Kockum}}, \bibinfo {author} {\bibfnamefont {A.}~\bibnamefont {Miranowicz}},
  \bibinfo {author} {\bibfnamefont {V.}~\bibnamefont {Macr\`{\i}}}, \bibinfo
  {author} {\bibfnamefont {S.}~\bibnamefont {Savasta}}, \ and\ \bibinfo
  {author} {\bibfnamefont {F.}~\bibnamefont {Nori}},\ }\bibfield  {title}
  {\enquote {\bibinfo {title} {Deterministic quantum nonlinear optics with
  single atoms and virtual photons},}\ }\href {\doibase
  10.1103/PhysRevA.95.063849} {\bibfield  {journal} {\bibinfo  {journal} {Phys.
  Rev. A}\ }\textbf {\bibinfo {volume} {95}},\ \bibinfo {pages} {063849}
  (\bibinfo {year} {2017}{\natexlab{a}})}\BibitemShut {NoStop}%
\bibitem [{\citenamefont {Kockum}\ \emph
  {et~al.}(2017{\natexlab{b}})\citenamefont {Kockum}, \citenamefont
  {Macr{\`\i}}, \citenamefont {Garziano}, \citenamefont {Savasta},\ and\
  \citenamefont {Nori}}]{Kockum2017b}%
  \BibitemOpen
  \bibfield  {author} {\bibinfo {author} {\bibfnamefont {A.~F.}\ \bibnamefont
  {Kockum}}, \bibinfo {author} {\bibfnamefont {V.}~\bibnamefont {Macr{\`\i}}},
  \bibinfo {author} {\bibfnamefont {L.}~\bibnamefont {Garziano}}, \bibinfo
  {author} {\bibfnamefont {S.}~\bibnamefont {Savasta}}, \ and\ \bibinfo
  {author} {\bibfnamefont {F.}~\bibnamefont {Nori}},\ }\bibfield  {title}
  {\enquote {\bibinfo {title} {Frequency conversion in ultrastrong cavity
  {QED}},}\ }\href
  {https://www.nature.com/articles/s41598-017-04225-3?WT.feed_name=subjects_quantum-physics}
  {\bibfield  {journal} {\bibinfo  {journal} {Sci. Rep.}\ }\textbf {\bibinfo
  {volume} {7}},\ \bibinfo {pages} {5313} (\bibinfo {year}
  {2017}{\natexlab{b}})}\BibitemShut {NoStop}%
\bibitem [{\citenamefont {Sackett}\ \emph {et~al.}(2000)\citenamefont
  {Sackett}, \citenamefont {Kielpinski}, \citenamefont {King}, \citenamefont
  {Langer}, \citenamefont {V.~Meyer}, \citenamefont {Myatt}, \citenamefont
  {Rowe}, \citenamefont {Turchette}, \citenamefont {Itano}, \citenamefont
  {Wineland},\ and\ \citenamefont {Monroe}}]{Sackett2000}%
  \BibitemOpen
  \bibfield  {author} {\bibinfo {author} {\bibfnamefont {C.~A.}\ \bibnamefont
  {Sackett}}, \bibinfo {author} {\bibfnamefont {D.}~\bibnamefont {Kielpinski}},
  \bibinfo {author} {\bibfnamefont {B.~E.}\ \bibnamefont {King}}, \bibinfo
  {author} {\bibfnamefont {C.}~\bibnamefont {Langer}}, \bibinfo {author}
  {\bibfnamefont {V.}~\bibnamefont {V.~Meyer}}, \bibinfo {author}
  {\bibfnamefont {C.~J.}\ \bibnamefont {Myatt}}, \bibinfo {author}
  {\bibfnamefont {M.}~\bibnamefont {Rowe}}, \bibinfo {author} {\bibfnamefont
  {Q.~A.}\ \bibnamefont {Turchette}}, \bibinfo {author} {\bibfnamefont {W.~M.}\
  \bibnamefont {Itano}}, \bibinfo {author} {\bibfnamefont {D.~J.}\ \bibnamefont
  {Wineland}}, \ and\ \bibinfo {author} {\bibfnamefont {C.}~\bibnamefont
  {Monroe}},\ }\bibfield  {title} {\enquote {\bibinfo {title} {Experimental
  entanglement of four particles},}\ }\href
  {https://www.nature.com/nature/journal/v404/n6775/full/404256a0.html}
  {\bibfield  {journal} {\bibinfo  {journal} {Nature}\ }\textbf {\bibinfo
  {volume} {404}},\ \bibinfo {pages} {256} (\bibinfo {year}
  {2000})}\BibitemShut {NoStop}%
\bibitem [{\citenamefont {Leibfried}\ \emph {et~al.}(2003)\citenamefont
  {Leibfried}, \citenamefont {DeMarco}, \citenamefont {Meyer}, \citenamefont
  {Lucas}, \citenamefont {Barrett}, \citenamefont {Britton}, \citenamefont
  {Itano}, \citenamefont {Jelenkovi\'{c}}, \citenamefont {Langer},
  \citenamefont {Rosenband},\ and\ \citenamefont {Wineland}}]{Leibfried2003}%
  \BibitemOpen
  \bibfield  {author} {\bibinfo {author} {\bibfnamefont {D.}~\bibnamefont
  {Leibfried}}, \bibinfo {author} {\bibfnamefont {B.}~\bibnamefont {DeMarco}},
  \bibinfo {author} {\bibfnamefont {V.}~\bibnamefont {Meyer}}, \bibinfo
  {author} {\bibfnamefont {D.}~\bibnamefont {Lucas}}, \bibinfo {author}
  {\bibfnamefont {M.}~\bibnamefont {Barrett}}, \bibinfo {author} {\bibfnamefont
  {J.}~\bibnamefont {Britton}}, \bibinfo {author} {\bibfnamefont {W.~M.}\
  \bibnamefont {Itano}}, \bibinfo {author} {\bibfnamefont {B.}~\bibnamefont
  {Jelenkovi\'{c}}}, \bibinfo {author} {\bibfnamefont {C.}~\bibnamefont
  {Langer}}, \bibinfo {author} {\bibfnamefont {T.}~\bibnamefont {Rosenband}}, \
  and\ \bibinfo {author} {\bibfnamefont {D.~J.}\ \bibnamefont {Wineland}},\
  }\bibfield  {title} {\enquote {\bibinfo {title} {Experimental demonstration
  of a robust, high-fidelity geometric two ion-qubit phase gate},}\ }\href
  {http://www.nature.com/nature/journal/v422/n6930/abs/nature01492.html}
  {\bibfield  {journal} {\bibinfo  {journal} {Nature}\ }\textbf {\bibinfo
  {volume} {422}},\ \bibinfo {pages} {412} (\bibinfo {year}
  {2003})}\BibitemShut {NoStop}%
\bibitem [{\citenamefont {Chen}\ \emph {et~al.}(2018)\citenamefont {Chen},
  \citenamefont {Shi}, \citenamefont {Song},\ and\ \citenamefont
  {Xia}}]{chen2018}%
  \BibitemOpen
  \bibfield  {author} {\bibinfo {author} {\bibfnamefont {Y.}~\bibnamefont
  {Chen}}, \bibinfo {author} {\bibfnamefont {Zhi-C.}\ \bibnamefont {Shi}},
  \bibinfo {author} {\bibfnamefont {J.}~\bibnamefont {Song}}, \ and\ \bibinfo
  {author} {\bibfnamefont {Y.}~\bibnamefont {Xia}},\ }\bibfield  {title}
  {\enquote {\bibinfo {title} {Invariant-based inverse engineering for
  fluctuation transfer between membranes in an optomechanical cavity system},}\
  }\href {\doibase https://doi.org/10.1103/PhysRevA.97.023841} {\bibfield
  {journal} {\bibinfo  {journal} {Physical Review A}\ }\textbf {\bibinfo
  {volume} {97}},\ \bibinfo {pages} {023841} (\bibinfo {year}
  {2018})}\BibitemShut {NoStop}%
\bibitem [{\citenamefont {Kardar}\ and\ \citenamefont
  {Golestanian}(1999)}]{Kardar1999}%
  \BibitemOpen
  \bibfield  {author} {\bibinfo {author} {\bibfnamefont {M.}~\bibnamefont
  {Kardar}}\ and\ \bibinfo {author} {\bibfnamefont {R.}~\bibnamefont
  {Golestanian}},\ }\bibfield  {title} {\enquote {\bibinfo {title} {The
  ``friction'' of vacuum, and other fluctuation-induced forces},}\ }\href
  {\doibase 10.1103/RevModPhys.71.1233} {\bibfield  {journal} {\bibinfo
  {journal} {Rev. Mod. Phys.}\ }\textbf {\bibinfo {volume} {71}},\ \bibinfo
  {pages} {1233} (\bibinfo {year} {1999})}\BibitemShut {NoStop}%
\bibitem [{\citenamefont {Dalvit}\ and\ \citenamefont
  {Maia~Neto}(2000)}]{Dalvit2000}%
  \BibitemOpen
  \bibfield  {author} {\bibinfo {author} {\bibfnamefont {D.~A.~R.}\
  \bibnamefont {Dalvit}}\ and\ \bibinfo {author} {\bibfnamefont {P.~A.}\
  \bibnamefont {Maia~Neto}},\ }\bibfield  {title} {\enquote {\bibinfo {title}
  {Decoherence via the dynamical {C}asimir effect},}\ }\href {\doibase
  10.1103/PhysRevLett.84.798} {\bibfield  {journal} {\bibinfo  {journal} {Phys.
  Rev. Lett.}\ }\textbf {\bibinfo {volume} {84}},\ \bibinfo {pages} {798}
  (\bibinfo {year} {2000})}\BibitemShut {NoStop}%
\bibitem [{\citenamefont {Jackel}\ and\ \citenamefont
  {Reynaud}(1992)}]{Jaekel1992a}%
  \BibitemOpen
  \bibfield  {author} {\bibinfo {author} {\bibfnamefont {M.~T.}\ \bibnamefont
  {Jackel}}\ and\ \bibinfo {author} {\bibfnamefont {S.}~\bibnamefont
  {Reynaud}},\ }\bibfield  {title} {\enquote {\bibinfo {title} {Fluctuations
  and dissipation for a mirror in vacuum},}\ }\href
  {http://iopscience.iop.org/article/10.1088/0954-8998/4/1/005} {\bibfield
  {journal} {\bibinfo  {journal} {Quantum Opt.}\ }\textbf {\bibinfo {volume}
  {4}},\ \bibinfo {pages} {39} (\bibinfo {year} {1992})}\BibitemShut {NoStop}%
\bibitem [{\citenamefont {Jaekel}\ and\ \citenamefont
  {Reynaud}(1992)}]{Jaekel1992b}%
  \BibitemOpen
  \bibfield  {author} {\bibinfo {author} {\bibfnamefont {M.~T.}\ \bibnamefont
  {Jaekel}}\ and\ \bibinfo {author} {\bibfnamefont {S.}~\bibnamefont
  {Reynaud}},\ }\bibfield  {title} {\enquote {\bibinfo {title} {Motional
  {C}asimir force},}\ }\href
  {https://jp1.journaldephysique.org/en/articles/jp1/abs/1992/02/jp1v2p149/jp1v2p149.html}
  {\bibfield  {journal} {\bibinfo  {journal} {J. Phys. I}\ }\textbf {\bibinfo
  {volume} {2}},\ \bibinfo {pages} {149} (\bibinfo {year} {1992})}\BibitemShut
  {NoStop}%
\bibitem [{\citenamefont {O'Connell}\ \emph {et~al.}(2010)\citenamefont
  {O'Connell}, \citenamefont {Hofheinz}, \citenamefont {Ansmann}, \citenamefont
  {Bialczak}, \citenamefont {Lenander}, \citenamefont {Lucero}, \citenamefont
  {Neeley}, \citenamefont {Sank}, \citenamefont {Wang}, \citenamefont {Weides}
  \emph {et~al.}}]{OConnell2010}%
  \BibitemOpen
  \bibfield  {author} {\bibinfo {author} {\bibfnamefont {A.~D.}\ \bibnamefont
  {O'Connell}}, \bibinfo {author} {\bibfnamefont {M.}~\bibnamefont {Hofheinz}},
  \bibinfo {author} {\bibfnamefont {M.}~\bibnamefont {Ansmann}}, \bibinfo
  {author} {\bibfnamefont {R.~C.}\ \bibnamefont {Bialczak}}, \bibinfo {author}
  {\bibfnamefont {M.}~\bibnamefont {Lenander}}, \bibinfo {author}
  {\bibfnamefont {E.}~\bibnamefont {Lucero}}, \bibinfo {author} {\bibfnamefont
  {M.}~\bibnamefont {Neeley}}, \bibinfo {author} {\bibfnamefont
  {D.}~\bibnamefont {Sank}}, \bibinfo {author} {\bibfnamefont {H.}~\bibnamefont
  {Wang}}, \bibinfo {author} {\bibfnamefont {M.}~\bibnamefont {Weides}},  \emph
  {et~al.},\ }\bibfield  {title} {\enquote {\bibinfo {title} {Quantum ground
  state and single-phonon control of a mechanical resonator},}\ }\href
  {\doibase 10.1038/nature08967} {\bibfield  {journal} {\bibinfo  {journal}
  {Nature}\ }\textbf {\bibinfo {volume} {464}},\ \bibinfo {pages} {697}
  (\bibinfo {year} {2010})}\BibitemShut {NoStop}%
\bibitem [{\citenamefont {Rouxinol}\ \emph {et~al.}(2016)\citenamefont
  {Rouxinol}, \citenamefont {Hao}, \citenamefont {Brito}, \citenamefont
  {Caldeira}, \citenamefont {Irish},\ and\ \citenamefont
  {LaHaye}}]{Rouxinol2016}%
  \BibitemOpen
  \bibfield  {author} {\bibinfo {author} {\bibfnamefont {F.}~\bibnamefont
  {Rouxinol}}, \bibinfo {author} {\bibfnamefont {Y.}~\bibnamefont {Hao}},
  \bibinfo {author} {\bibfnamefont {F.}~\bibnamefont {Brito}}, \bibinfo
  {author} {\bibfnamefont {A.~O.}\ \bibnamefont {Caldeira}}, \bibinfo {author}
  {\bibfnamefont {E.~K.}\ \bibnamefont {Irish}}, \ and\ \bibinfo {author}
  {\bibfnamefont {M.~D.}\ \bibnamefont {LaHaye}},\ }\bibfield  {title}
  {\enquote {\bibinfo {title} {Measurements of nanoresonator-qubit interactions
  in a hybrid quantum electromechanical system},}\ }\href {\doibase
  10.1088/0957-4484/27/36/364003} {\bibfield  {journal} {\bibinfo  {journal}
  {Nanotechnology}\ }\textbf {\bibinfo {volume} {27}},\ \bibinfo {pages}
  {364003} (\bibinfo {year} {2016})}\BibitemShut {NoStop}%
\bibitem [{\citenamefont {Ockeloen-Korppi}\ \emph {et~al.}(2017)\citenamefont
  {Ockeloen-Korppi}, \citenamefont {Damskagg}, \citenamefont {Pirkkalainen},
  \citenamefont {Clerk}, \citenamefont {Massel}, \citenamefont {Woolley},\ and\
  \citenamefont {Sillanpaa}}]{Ockeloen2017}%
  \BibitemOpen
  \bibfield  {author} {\bibinfo {author} {\bibfnamefont {C.~F.}\ \bibnamefont
  {Ockeloen-Korppi}}, \bibinfo {author} {\bibfnamefont {E.}~\bibnamefont
  {Damskagg}}, \bibinfo {author} {\bibfnamefont {J.~M.}\ \bibnamefont
  {Pirkkalainen}}, \bibinfo {author} {\bibfnamefont {A.~A.}\ \bibnamefont
  {Clerk}}, \bibinfo {author} {\bibfnamefont {F.}~\bibnamefont {Massel}},
  \bibinfo {author} {\bibfnamefont {M.~J.}\ \bibnamefont {Woolley}}, \ and\
  \bibinfo {author} {\bibfnamefont {M.~A.}\ \bibnamefont {Sillanpaa}},\
  }\bibfield  {title} {\enquote {\bibinfo {title} {Entangled massive mechanical
  oscillators},}\ }\href {https://arxiv.org/abs/1711.01640} {\bibfield
  {journal} {\bibinfo  {journal} {arXiv:1711.01640}\ } (\bibinfo {year}
  {2017})}\BibitemShut {NoStop}%
\bibitem [{\citenamefont {Law}(1995)}]{Law1995}%
  \BibitemOpen
  \bibfield  {author} {\bibinfo {author} {\bibfnamefont {C.~K.}\ \bibnamefont
  {Law}},\ }\bibfield  {title} {\enquote {\bibinfo {title} {Interaction between
  a moving mirror and radiation pressure: A {H}amiltonian formulation},}\
  }\href {\doibase 10.1103/PhysRevA.51.2537} {\bibfield  {journal} {\bibinfo
  {journal} {Phys. Rev. A}\ }\textbf {\bibinfo {volume} {51}},\ \bibinfo
  {pages} {2537} (\bibinfo {year} {1995})}\BibitemShut {NoStop}%
\bibitem [{\citenamefont {Butera}\ and\ \citenamefont
  {Passante}(2013)}]{Butera2013}%
  \BibitemOpen
  \bibfield  {author} {\bibinfo {author} {\bibfnamefont {S.}~\bibnamefont
  {Butera}}\ and\ \bibinfo {author} {\bibfnamefont {R.}~\bibnamefont
  {Passante}},\ }\bibfield  {title} {\enquote {\bibinfo {title} {Field
  fluctuations in a one-dimensional cavity with a mobile wall},}\ }\href
  {\doibase 10.1103/PhysRevLett.111.060403} {\bibfield  {journal} {\bibinfo
  {journal} {Phys. Rev. Lett.}\ }\textbf {\bibinfo {volume} {111}},\ \bibinfo
  {pages} {060403} (\bibinfo {year} {2013})}\BibitemShut {NoStop}%
\bibitem [{\citenamefont {Macr\`{\i}}\ \emph {et~al.}(2018)\citenamefont
  {Macr\`{\i}}, \citenamefont {Ridolfo}, \citenamefont {Di~Stefano},
  \citenamefont {Kockum}, \citenamefont {Nori},\ and\ \citenamefont
  {Savasta}}]{Macri2017}%
  \BibitemOpen
  \bibfield  {author} {\bibinfo {author} {\bibfnamefont {V.}~\bibnamefont
  {Macr\`{\i}}}, \bibinfo {author} {\bibfnamefont {A.}~\bibnamefont {Ridolfo}},
  \bibinfo {author} {\bibfnamefont {O.}~\bibnamefont {Di~Stefano}}, \bibinfo
  {author} {\bibfnamefont {A.~F.}\ \bibnamefont {Kockum}}, \bibinfo {author}
  {\bibfnamefont {F.}~\bibnamefont {Nori}}, \ and\ \bibinfo {author}
  {\bibfnamefont {S.}~\bibnamefont {Savasta}},\ }\bibfield  {title} {\enquote
  {\bibinfo {title} {Nonperturbative dynamical casimir effect in optomechanical
  systems: Vacuum casimir-rabi splittings},}\ }\href {\doibase
  10.1103/PhysRevX.8.011031} {\bibfield  {journal} {\bibinfo  {journal} {Phys.
  Rev. X}\ }\textbf {\bibinfo {volume} {8}},\ \bibinfo {pages} {011031}
  (\bibinfo {year} {2018})}\BibitemShut {NoStop}%
\bibitem [{\citenamefont {Sala}\ and\ \citenamefont
  {Tufarelli}(2017)}]{Sala2017}%
  \BibitemOpen
  \bibfield  {author} {\bibinfo {author} {\bibfnamefont {K.}~\bibnamefont
  {Sala}}\ and\ \bibinfo {author} {\bibfnamefont {T.}~\bibnamefont
  {Tufarelli}},\ }\bibfield  {title} {\enquote {\bibinfo {title} {Exploring
  corrections to the optomechanical {H}amiltonian},}\ }\href
  {https://128.84.21.199/abs/1711.06688} {\bibfield  {journal} {\bibinfo
  {journal} {arXiv:1711.06688}\ } (\bibinfo {year} {2017})}\BibitemShut
  {NoStop}%
\bibitem [{\citenamefont {Armata}\ \emph {et~al.}(2017)\citenamefont {Armata},
  \citenamefont {Kim}, \citenamefont {Butera}, \citenamefont {Rizzuto},\ and\
  \citenamefont {Passante}}]{Armata2017}%
  \BibitemOpen
  \bibfield  {author} {\bibinfo {author} {\bibfnamefont {F.}~\bibnamefont
  {Armata}}, \bibinfo {author} {\bibfnamefont {M.~S.}\ \bibnamefont {Kim}},
  \bibinfo {author} {\bibfnamefont {S.}~\bibnamefont {Butera}}, \bibinfo
  {author} {\bibfnamefont {L.}~\bibnamefont {Rizzuto}}, \ and\ \bibinfo
  {author} {\bibfnamefont {R.}~\bibnamefont {Passante}},\ }\bibfield  {title}
  {\enquote {\bibinfo {title} {Nonequilibrium dressing in a cavity with a
  movable reflecting mirror},}\ }\href {\doibase 10.1103/PhysRevD.96.045007}
  {\bibfield  {journal} {\bibinfo  {journal} {Phys. Rev. D}\ }\textbf {\bibinfo
  {volume} {96}},\ \bibinfo {pages} {045007} (\bibinfo {year}
  {2017})}\BibitemShut {NoStop}%
\bibitem [{\citenamefont {Aspelmeyer}\ \emph {et~al.}(2014)\citenamefont
  {Aspelmeyer}, \citenamefont {Kippenberg},\ and\ \citenamefont
  {Marquardt}}]{Aspelmeyer2014}%
  \BibitemOpen
  \bibfield  {author} {\bibinfo {author} {\bibfnamefont {M.}~\bibnamefont
  {Aspelmeyer}}, \bibinfo {author} {\bibfnamefont {T.~J.}\ \bibnamefont
  {Kippenberg}}, \ and\ \bibinfo {author} {\bibfnamefont {F.}~\bibnamefont
  {Marquardt}},\ }\bibfield  {title} {\enquote {\bibinfo {title} {Cavity
  optomechanics},}\ }\href {\doibase 10.1103/RevModPhys.86.1391} {\bibfield
  {journal} {\bibinfo  {journal} {Rev. Mod. Phys.}\ }\textbf {\bibinfo {volume}
  {86}},\ \bibinfo {pages} {1391} (\bibinfo {year} {2014})}\BibitemShut
  {NoStop}%
\bibitem [{\citenamefont {Moore}(1970)}]{moore1970}%
  \BibitemOpen
  \bibfield  {author} {\bibinfo {author} {\bibfnamefont {G.~T.}\ \bibnamefont
  {Moore}},\ }\bibfield  {title} {\enquote {\bibinfo {title} {{Quantum Theory
  of the Electromagnetic Field in a Variable-Length One-Dimensional Cavity}},}\
  }\href {\doibase 10.1063/1.1665432} {\bibfield  {journal} {\bibinfo
  {journal} {J. Math. Phys.}\ }\textbf {\bibinfo {volume} {11}},\ \bibinfo
  {pages} {2679} (\bibinfo {year} {1970})}\BibitemShut {NoStop}%
\bibitem [{\citenamefont {Nation}\ \emph {et~al.}(2012)\citenamefont {Nation},
  \citenamefont {Johansson}, \citenamefont {Blencowe},\ and\ \citenamefont
  {Nori}}]{Nation2012}%
  \BibitemOpen
  \bibfield  {author} {\bibinfo {author} {\bibfnamefont {P.~D.}\ \bibnamefont
  {Nation}}, \bibinfo {author} {\bibfnamefont {J.~R.}\ \bibnamefont
  {Johansson}}, \bibinfo {author} {\bibfnamefont {M.~P.}\ \bibnamefont
  {Blencowe}}, \ and\ \bibinfo {author} {\bibfnamefont {Franco}\ \bibnamefont
  {Nori}},\ }\bibfield  {title} {\enquote {\bibinfo {title} {Colloquium:
  {S}timulating uncertainty:{ A}mplifying the quantum vacuum with
  superconducting circuits},}\ }\href {\doibase 10.1103/RevModPhys.84.1}
  {\bibfield  {journal} {\bibinfo  {journal} {Rev. Mod. Phys.}\ }\textbf
  {\bibinfo {volume} {84}},\ \bibinfo {pages} {1--24} (\bibinfo {year}
  {2012})}\BibitemShut {NoStop}%
\bibitem [{\citenamefont {Johansson}\ \emph {et~al.}(2009)\citenamefont
  {Johansson}, \citenamefont {Johansson}, \citenamefont {Wilson},\ and\
  \citenamefont {Nori}}]{Johansson2009}%
  \BibitemOpen
  \bibfield  {author} {\bibinfo {author} {\bibfnamefont {J.~R.}\ \bibnamefont
  {Johansson}}, \bibinfo {author} {\bibfnamefont {G.}~\bibnamefont
  {Johansson}}, \bibinfo {author} {\bibfnamefont {C.~M.}\ \bibnamefont
  {Wilson}}, \ and\ \bibinfo {author} {\bibfnamefont {F.}~\bibnamefont
  {Nori}},\ }\bibfield  {title} {\enquote {\bibinfo {title} {{Dynamical
  {C}asimir Effect in a Superconducting Coplanar Waveguide}},}\ }\href
  {\doibase 10.1103/PhysRevLett.103.147003} {\bibfield  {journal} {\bibinfo
  {journal} {Phys. Rev. Lett.}\ }\textbf {\bibinfo {volume} {103}},\ \bibinfo
  {pages} {147003} (\bibinfo {year} {2009})}\BibitemShut {NoStop}%
\bibitem [{\citenamefont {Johansson}\ \emph {et~al.}(2010)\citenamefont
  {Johansson}, \citenamefont {Johansson}, \citenamefont {Wilson},\ and\
  \citenamefont {Nori}}]{Johansson2010}%
  \BibitemOpen
  \bibfield  {author} {\bibinfo {author} {\bibfnamefont {J.~R.}\ \bibnamefont
  {Johansson}}, \bibinfo {author} {\bibfnamefont {G.}~\bibnamefont
  {Johansson}}, \bibinfo {author} {\bibfnamefont {C.~M.}\ \bibnamefont
  {Wilson}}, \ and\ \bibinfo {author} {\bibfnamefont {F.}~\bibnamefont
  {Nori}},\ }\bibfield  {title} {\enquote {\bibinfo {title} {{Dynamical
  {C}asimir effect in superconducting microwave circuits}},}\ }\href {\doibase
  10.1103/PhysRevA.82.052509} {\bibfield  {journal} {\bibinfo  {journal} {Phys.
  Rev. A}\ }\textbf {\bibinfo {volume} {82}},\ \bibinfo {pages} {52509}
  (\bibinfo {year} {2010})}\BibitemShut {NoStop}%
\bibitem [{\citenamefont {Wilson}\ \emph {et~al.}(2011)\citenamefont {Wilson},
  \citenamefont {Johansson}, \citenamefont {Pourkabirian}, \citenamefont
  {Simoen}, \citenamefont {Johansson}, \citenamefont {Duty}, \citenamefont
  {Nori},\ and\ \citenamefont {Delsing}}]{Wilson2011}%
  \BibitemOpen
  \bibfield  {author} {\bibinfo {author} {\bibfnamefont {C.~M.}\ \bibnamefont
  {Wilson}}, \bibinfo {author} {\bibfnamefont {G.}~\bibnamefont {Johansson}},
  \bibinfo {author} {\bibfnamefont {A.}~\bibnamefont {Pourkabirian}}, \bibinfo
  {author} {\bibfnamefont {M.}~\bibnamefont {Simoen}}, \bibinfo {author}
  {\bibfnamefont {J.~R.}\ \bibnamefont {Johansson}}, \bibinfo {author}
  {\bibfnamefont {T.}~\bibnamefont {Duty}}, \bibinfo {author} {\bibfnamefont
  {F.}~\bibnamefont {Nori}}, \ and\ \bibinfo {author} {\bibfnamefont
  {P.}~\bibnamefont {Delsing}},\ }\bibfield  {title} {\enquote {\bibinfo
  {title} {{Observation of the dynamical {C}asimir effect in a superconducting
  circuit}},}\ }\href {\doibase 10.1038/nature10561} {\bibfield  {journal}
  {\bibinfo  {journal} {Nature}\ }\textbf {\bibinfo {volume} {479}},\ \bibinfo
  {pages} {376} (\bibinfo {year} {2011})}\BibitemShut {NoStop}%
\bibitem [{\citenamefont {Johansson}\ \emph {et~al.}(2013)\citenamefont
  {Johansson}, \citenamefont {Johansson}, \citenamefont {Wilson}, \citenamefont
  {Delsing},\ and\ \citenamefont {Nori}}]{Johansson2013}%
  \BibitemOpen
  \bibfield  {author} {\bibinfo {author} {\bibfnamefont {J.~R.}\ \bibnamefont
  {Johansson}}, \bibinfo {author} {\bibfnamefont {G.}~\bibnamefont
  {Johansson}}, \bibinfo {author} {\bibfnamefont {C.~M.}\ \bibnamefont
  {Wilson}}, \bibinfo {author} {\bibfnamefont {P.}~\bibnamefont {Delsing}}, \
  and\ \bibinfo {author} {\bibfnamefont {F.}~\bibnamefont {Nori}},\ }\bibfield
  {title} {\enquote {\bibinfo {title} {{Nonclassical microwave radiation from
  the dynamical {C}asimir effect}},}\ }\href {\doibase
  10.1103/PhysRevA.87.043804} {\bibfield  {journal} {\bibinfo  {journal} {Phys.
  Rev. A}\ }\textbf {\bibinfo {volume} {87}},\ \bibinfo {pages} {043804}
  (\bibinfo {year} {2013})}\BibitemShut {NoStop}%
\bibitem [{\citenamefont {Felicetti}\ \emph {et~al.}(2014)\citenamefont
  {Felicetti}, \citenamefont {Sanz}, \citenamefont {Lamata}, \citenamefont
  {Romero}, \citenamefont {Johansson}, \citenamefont {Delsing},\ and\
  \citenamefont {Solano}}]{felicetti2014}%
  \BibitemOpen
  \bibfield  {author} {\bibinfo {author} {\bibfnamefont {S.}~\bibnamefont
  {Felicetti}}, \bibinfo {author} {\bibfnamefont {M.}~\bibnamefont {Sanz}},
  \bibinfo {author} {\bibfnamefont {L.}~\bibnamefont {Lamata}}, \bibinfo
  {author} {\bibfnamefont {G.}~\bibnamefont {Romero}}, \bibinfo {author}
  {\bibfnamefont {G.}~\bibnamefont {Johansson}}, \bibinfo {author}
  {\bibfnamefont {P.}~\bibnamefont {Delsing}}, \ and\ \bibinfo {author}
  {\bibfnamefont {E.}~\bibnamefont {Solano}},\ }\bibfield  {title} {\enquote
  {\bibinfo {title} {Dynamical {C}asimir effect entangles artificial atoms},}\
  }\href {\doibase 10.1103/physrevlett.113.093602} {\bibfield  {journal}
  {\bibinfo  {journal} {Phys. Rev. Lett.}\ }\textbf {\bibinfo {volume} {113}},\
  \bibinfo {pages} {093602} (\bibinfo {year} {2014})}\BibitemShut {NoStop}%
\bibitem [{\citenamefont {Stassi}\ \emph {et~al.}(2015)\citenamefont {Stassi},
  \citenamefont {De~Liberato}, \citenamefont {Garziano}, \citenamefont
  {Spagnolo},\ and\ \citenamefont {Savasta}}]{stassi2015}%
  \BibitemOpen
  \bibfield  {author} {\bibinfo {author} {\bibfnamefont {R.}~\bibnamefont
  {Stassi}}, \bibinfo {author} {\bibfnamefont {S.}~\bibnamefont {De~Liberato}},
  \bibinfo {author} {\bibfnamefont {L.}~\bibnamefont {Garziano}}, \bibinfo
  {author} {\bibfnamefont {B.}~\bibnamefont {Spagnolo}}, \ and\ \bibinfo
  {author} {\bibfnamefont {S.}~\bibnamefont {Savasta}},\ }\bibfield  {title}
  {\enquote {\bibinfo {title} {Quantum control and long-range quantum
  correlations in dynamical {C}asimir arrays},}\ }\href {\doibase
  10.1103/physreva.92.013830} {\bibfield  {journal} {\bibinfo  {journal} {Phys.
  Rev. A}\ }\textbf {\bibinfo {volume} {92}},\ \bibinfo {pages} {013830}
  (\bibinfo {year} {2015})}\BibitemShut {NoStop}%
\bibitem [{\citenamefont {Rossatto}\ \emph {et~al.}(2016)\citenamefont
  {Rossatto}, \citenamefont {Felicetti}, \citenamefont {Eneriz}, \citenamefont
  {Rico}, \citenamefont {Sanz},\ and\ \citenamefont {Solano}}]{rossatto2016}%
  \BibitemOpen
  \bibfield  {author} {\bibinfo {author} {\bibfnamefont {D.~Z.}\ \bibnamefont
  {Rossatto}}, \bibinfo {author} {\bibfnamefont {S.}~\bibnamefont {Felicetti}},
  \bibinfo {author} {\bibfnamefont {H.}~\bibnamefont {Eneriz}}, \bibinfo
  {author} {\bibfnamefont {E.}~\bibnamefont {Rico}}, \bibinfo {author}
  {\bibfnamefont {M.}~\bibnamefont {Sanz}}, \ and\ \bibinfo {author}
  {\bibfnamefont {E.}~\bibnamefont {Solano}},\ }\bibfield  {title} {\enquote
  {\bibinfo {title} {Entangling polaritons via dynamical {C}asimir effect in
  circuit quantum electrodynamics},}\ }\href {\doibase
  10.1103/physrevb.93.094514} {\bibfield  {journal} {\bibinfo  {journal} {Phys.
  Rev. B}\ }\textbf {\bibinfo {volume} {93}},\ \bibinfo {pages} {094514}
  (\bibinfo {year} {2016})}\BibitemShut {NoStop}%
\bibitem [{\citenamefont {Heikkil\"a}\ \emph {et~al.}(2014)\citenamefont
  {Heikkil\"a}, \citenamefont {Massel}, \citenamefont {Tuorila}, \citenamefont
  {Khan},\ and\ \citenamefont {Sillanp\"a\"a}}]{Heikkila2014}%
  \BibitemOpen
  \bibfield  {author} {\bibinfo {author} {\bibfnamefont {T.~T.}\ \bibnamefont
  {Heikkil\"a}}, \bibinfo {author} {\bibfnamefont {F.}~\bibnamefont {Massel}},
  \bibinfo {author} {\bibfnamefont {J.}~\bibnamefont {Tuorila}}, \bibinfo
  {author} {\bibfnamefont {R.}~\bibnamefont {Khan}}, \ and\ \bibinfo {author}
  {\bibfnamefont {M.~A.}\ \bibnamefont {Sillanp\"a\"a}},\ }\bibfield  {title}
  {\enquote {\bibinfo {title} {Enhancing optomechanical coupling via the
  {J}osephson effect},}\ }\href {\doibase 10.1103/PhysRevLett.112.203603}
  {\bibfield  {journal} {\bibinfo  {journal} {Phys. Rev. Lett.}\ }\textbf
  {\bibinfo {volume} {112}},\ \bibinfo {pages} {203603} (\bibinfo {year}
  {2014})}\BibitemShut {NoStop}%
\bibitem [{\citenamefont {Beaudoin}\ \emph {et~al.}(2011)\citenamefont
  {Beaudoin}, \citenamefont {Gambetta},\ and\ \citenamefont
  {Blais}}]{Beaudoin2011}%
  \BibitemOpen
  \bibfield  {author} {\bibinfo {author} {\bibfnamefont {F.}~\bibnamefont
  {Beaudoin}}, \bibinfo {author} {\bibfnamefont {J.~M.}\ \bibnamefont
  {Gambetta}}, \ and\ \bibinfo {author} {\bibfnamefont {A.}~\bibnamefont
  {Blais}},\ }\bibfield  {title} {\enquote {\bibinfo {title} {{Dissipation and
  ultrastrong coupling in circuit QED}},}\ }\href {\doibase
  10.1103/PhysRevA.84.043832} {\bibfield  {journal} {\bibinfo  {journal} {Phys.
  Rev. A}\ }\textbf {\bibinfo {volume} {84}},\ \bibinfo {pages} {043832}
  (\bibinfo {year} {2011})}\BibitemShut {NoStop}%
\bibitem [{\citenamefont {Hu}\ \emph {et~al.}(2015)\citenamefont {Hu},
  \citenamefont {Huang}, \citenamefont {Liao}, \citenamefont {Tian},\ and\
  \citenamefont {Goan}}]{Hu2015}%
  \BibitemOpen
  \bibfield  {author} {\bibinfo {author} {\bibfnamefont {D.}~\bibnamefont
  {Hu}}, \bibinfo {author} {\bibfnamefont {S.-Y.}\ \bibnamefont {Huang}},
  \bibinfo {author} {\bibfnamefont {J.-Q.}\ \bibnamefont {Liao}}, \bibinfo
  {author} {\bibfnamefont {L.}~\bibnamefont {Tian}}, \ and\ \bibinfo {author}
  {\bibfnamefont {H.-S.}\ \bibnamefont {Goan}},\ }\bibfield  {title} {\enquote
  {\bibinfo {title} {Quantum coherence in ultrastrong optomechanics},}\ }\href
  {\doibase 10.1103/PhysRevA.91.013812} {\bibfield  {journal} {\bibinfo
  {journal} {Phys. Rev. A}\ }\textbf {\bibinfo {volume} {91}},\ \bibinfo
  {pages} {013812} (\bibinfo {year} {2015})}\BibitemShut {NoStop}%
\bibitem [{\citenamefont {Breuer}\ and\ \citenamefont
  {Petruccione}(2002)}]{Breuer2002}%
  \BibitemOpen
  \bibfield  {author} {\bibinfo {author} {\bibfnamefont {H.-P.}\ \bibnamefont
  {Breuer}}\ and\ \bibinfo {author} {\bibfnamefont {F.}~\bibnamefont
  {Petruccione}},\ }\href@noop {} {\emph {\bibinfo {title} {{The Theory of Open
  Quantum Systems}}}}\ (\bibinfo  {publisher} {Oxford University Press},\
  \bibinfo {year} {2002})\BibitemShut {NoStop}%
\bibitem [{\citenamefont {Ma}\ and\ \citenamefont {Law}(2015)}]{Ma2015}%
  \BibitemOpen
  \bibfield  {author} {\bibinfo {author} {\bibfnamefont {K.~K.~W.}\
  \bibnamefont {Ma}}\ and\ \bibinfo {author} {\bibfnamefont {C.~K.}\
  \bibnamefont {Law}},\ }\bibfield  {title} {\enquote {\bibinfo {title}
  {Three-photon resonance and adiabatic passage in the large-detuning {R}abi
  model},}\ }\href {\doibase 10.1103/PhysRevA.92.023842} {\bibfield  {journal}
  {\bibinfo  {journal} {Phys. Rev. A}\ }\textbf {\bibinfo {volume} {92}},\
  \bibinfo {pages} {023842} (\bibinfo {year} {2015})}\BibitemShut {NoStop}%
\bibitem [{\citenamefont {Garziano}\ \emph
  {et~al.}(2015{\natexlab{a}})\citenamefont {Garziano}, \citenamefont {Stassi},
  \citenamefont {Macr\'{\i}}, \citenamefont {Savasta},\ and\ \citenamefont
  {Di~Stefano}}]{Garziano2015b}%
  \BibitemOpen
  \bibfield  {author} {\bibinfo {author} {\bibfnamefont {L.}~\bibnamefont
  {Garziano}}, \bibinfo {author} {\bibfnamefont {R.}~\bibnamefont {Stassi}},
  \bibinfo {author} {\bibfnamefont {V.}~\bibnamefont {Macr\'{\i}}}, \bibinfo
  {author} {\bibfnamefont {S.}~\bibnamefont {Savasta}}, \ and\ \bibinfo
  {author} {\bibfnamefont {O.}~\bibnamefont {Di~Stefano}},\ }\bibfield  {title}
  {\enquote {\bibinfo {title} {Single-step arbitrary control of mechanical
  quantum states in ultrastrong optomechanics},}\ }\href {\doibase
  10.1103/PhysRevA.91.023809} {\bibfield  {journal} {\bibinfo  {journal} {Phys.
  Rev. A}\ }\textbf {\bibinfo {volume} {91}},\ \bibinfo {pages} {023809}
  (\bibinfo {year} {2015}{\natexlab{a}})}\BibitemShut {NoStop}%
\bibitem [{\citenamefont {Macr\'{\i}}\ \emph {et~al.}(2016)\citenamefont
  {Macr\'{\i}}, \citenamefont {Garziano}, \citenamefont {Ridolfo},
  \citenamefont {Di~Stefano},\ and\ \citenamefont {Savasta}}]{Macri2016}%
  \BibitemOpen
  \bibfield  {author} {\bibinfo {author} {\bibfnamefont {V.}~\bibnamefont
  {Macr\'{\i}}}, \bibinfo {author} {\bibfnamefont {L.}~\bibnamefont
  {Garziano}}, \bibinfo {author} {\bibfnamefont {A.}~\bibnamefont {Ridolfo}},
  \bibinfo {author} {\bibfnamefont {O.}~\bibnamefont {Di~Stefano}}, \ and\
  \bibinfo {author} {\bibfnamefont {S.}~\bibnamefont {Savasta}},\ }\bibfield
  {title} {\enquote {\bibinfo {title} {Deterministic synthesis of mechanical
  {NOON} states in ultrastrong optomechanics},}\ }\href {\doibase
  10.1103/PhysRevA.94.013817} {\bibfield  {journal} {\bibinfo  {journal} {Phys.
  Rev. A}\ }\textbf {\bibinfo {volume} {94}},\ \bibinfo {pages} {013817}
  (\bibinfo {year} {2016})}\BibitemShut {NoStop}%
\bibitem [{\citenamefont {Gamel}\ and\ \citenamefont
  {James}(2010)}]{Gamel2010}%
  \BibitemOpen
  \bibfield  {author} {\bibinfo {author} {\bibfnamefont {O.}~\bibnamefont
  {Gamel}}\ and\ \bibinfo {author} {\bibfnamefont {D.~F.~V.}\ \bibnamefont
  {James}},\ }\bibfield  {title} {\enquote {\bibinfo {title} {Time-averaged
  quantum dynamics and the validity of the effective {H}amiltonian model},}\
  }\href {\doibase 10.1103/PhysRevA.82.052106} {\bibfield  {journal} {\bibinfo
  {journal} {Phys. Rev. A}\ }\textbf {\bibinfo {volume} {82}},\ \bibinfo
  {pages} {052106} (\bibinfo {year} {2010})}\BibitemShut {NoStop}%
\bibitem [{\citenamefont {Shao}\ \emph {et~al.}(2017)\citenamefont {Shao},
  \citenamefont {Wu},\ and\ \citenamefont {Feng}}]{Shao2017}%
  \BibitemOpen
  \bibfield  {author} {\bibinfo {author} {\bibfnamefont {W.}~\bibnamefont
  {Shao}}, \bibinfo {author} {\bibfnamefont {C.}~\bibnamefont {Wu}}, \ and\
  \bibinfo {author} {\bibfnamefont {X.-L.}\ \bibnamefont {Feng}},\ }\bibfield
  {title} {\enquote {\bibinfo {title} {Generalized {J}ames' effective
  {H}amiltonian method},}\ }\href {\doibase 10.1103/physreva.95.032124}
  {\bibfield  {journal} {\bibinfo  {journal} {Phys. Rev. A}\ }\textbf {\bibinfo
  {volume} {95}},\ \bibinfo {pages} {032124} (\bibinfo {year}
  {2017})}\BibitemShut {NoStop}%
\bibitem [{\citenamefont {Settineri}\ \emph {et~al.}(2018)\citenamefont
  {Settineri}, \citenamefont {Macr{\`{i}}}, \citenamefont {Ridolfo},
  \citenamefont {Di~Stefano}, \citenamefont {Kockum}, \citenamefont {Nori},\
  and\ \citenamefont {Savasta}}]{Settineri2018}%
  \BibitemOpen
  \bibfield  {author} {\bibinfo {author} {\bibfnamefont {A.}~\bibnamefont
  {Settineri}}, \bibinfo {author} {\bibfnamefont {V.}~\bibnamefont
  {Macr{\`{i}}}}, \bibinfo {author} {\bibfnamefont {A.}~\bibnamefont
  {Ridolfo}}, \bibinfo {author} {\bibfnamefont {O.}~\bibnamefont {Di~Stefano}},
  \bibinfo {author} {\bibfnamefont {A.~F.}\ \bibnamefont {Kockum}}, \bibinfo
  {author} {\bibfnamefont {F.}~\bibnamefont {Nori}}, \ and\ \bibinfo {author}
  {\bibfnamefont {S.}~\bibnamefont {Savasta}},\ }\bibfield  {title} {\enquote
  {\bibinfo {title} {Dissipation and thermal noise in hybrid quantum systems in
  the ultrastrong coupling regime},}\ }\href {https://arxiv.org/abs/1807.06348}
  {\bibfield  {journal} {\bibinfo  {journal} {arXiv:1807.06348}\ } (\bibinfo
  {year} {2018})}\BibitemShut {NoStop}%
\bibitem [{\citenamefont {Ridolfo}\ \emph {et~al.}(2012)\citenamefont
  {Ridolfo}, \citenamefont {Leib}, \citenamefont {Savasta},\ and\ \citenamefont
  {Hartmann}}]{Ridolfo2012}%
  \BibitemOpen
  \bibfield  {author} {\bibinfo {author} {\bibfnamefont {A.}~\bibnamefont
  {Ridolfo}}, \bibinfo {author} {\bibfnamefont {M.}~\bibnamefont {Leib}},
  \bibinfo {author} {\bibfnamefont {S.}~\bibnamefont {Savasta}}, \ and\
  \bibinfo {author} {\bibfnamefont {M.~J.}\ \bibnamefont {Hartmann}},\
  }\bibfield  {title} {\enquote {\bibinfo {title} {{Photon Blockade in the
  Ultrastrong Coupling Regime}},}\ }\href {\doibase
  10.1103/PhysRevLett.109.193602} {\bibfield  {journal} {\bibinfo  {journal}
  {Phys. Rev. Lett.}\ }\textbf {\bibinfo {volume} {109}},\ \bibinfo {pages}
  {193602} (\bibinfo {year} {2012})}\BibitemShut {NoStop}%
\bibitem [{\citenamefont {Garziano}\ \emph
  {et~al.}(2015{\natexlab{b}})\citenamefont {Garziano}, \citenamefont {Stassi},
  \citenamefont {Macr{\`{\i}}}, \citenamefont {Kockum}, \citenamefont
  {Savasta},\ and\ \citenamefont {Nori}}]{Garziano2015}%
  \BibitemOpen
  \bibfield  {author} {\bibinfo {author} {\bibfnamefont {L.}~\bibnamefont
  {Garziano}}, \bibinfo {author} {\bibfnamefont {R.}~\bibnamefont {Stassi}},
  \bibinfo {author} {\bibfnamefont {V.}~\bibnamefont {Macr{\`{\i}}}}, \bibinfo
  {author} {\bibfnamefont {A.~F.}\ \bibnamefont {Kockum}}, \bibinfo {author}
  {\bibfnamefont {S.}~\bibnamefont {Savasta}}, \ and\ \bibinfo {author}
  {\bibfnamefont {F.}~\bibnamefont {Nori}},\ }\bibfield  {title} {\enquote
  {\bibinfo {title} {{Multiphoton quantum Rabi oscillations in ultrastrong
  cavity QED}},}\ }\href {\doibase 10.1103/PhysRevA.92.063830} {\bibfield
  {journal} {\bibinfo  {journal} {Phys. Rev. A}\ }\textbf {\bibinfo {volume}
  {92}},\ \bibinfo {pages} {063830} (\bibinfo {year}
  {2015}{\natexlab{b}})}\BibitemShut {NoStop}%
\bibitem [{\citenamefont {Garziano}\ \emph {et~al.}(2016)\citenamefont
  {Garziano}, \citenamefont {Stassi}, \citenamefont {Macr{\`{\i}}},
  \citenamefont {{Di Stefano}}, \citenamefont {Nori},\ and\ \citenamefont
  {Savasta}}]{Garziano2016}%
  \BibitemOpen
  \bibfield  {author} {\bibinfo {author} {\bibfnamefont {L.}~\bibnamefont
  {Garziano}}, \bibinfo {author} {\bibfnamefont {R.}~\bibnamefont {Stassi}},
  \bibinfo {author} {\bibfnamefont {V.}~\bibnamefont {Macr{\`{\i}}}}, \bibinfo
  {author} {\bibfnamefont {O.}~\bibnamefont {{Di Stefano}}}, \bibinfo {author}
  {\bibfnamefont {F.}~\bibnamefont {Nori}}, \ and\ \bibinfo {author}
  {\bibfnamefont {S.}~\bibnamefont {Savasta}},\ }\bibfield  {title} {\enquote
  {\bibinfo {title} {{One Photon Can Simultaneously Excite Two or More
  Atoms}},}\ }\href {\doibase 10.1103/PhysRevLett.117.043601} {\bibfield
  {journal} {\bibinfo  {journal} {Phys. Rev. Lett.}\ }\textbf {\bibinfo
  {volume} {117}},\ \bibinfo {pages} {043601} (\bibinfo {year}
  {2016})}\BibitemShut {NoStop}%
\bibitem [{\citenamefont {Stannigel}\ \emph {et~al.}(2012)\citenamefont
  {Stannigel}, \citenamefont {Komar}, \citenamefont {Habraken}, \citenamefont
  {Bennett}, \citenamefont {Lukin}, \citenamefont {Zoller},\ and\ \citenamefont
  {Rabl}}]{Stannigel2012}%
  \BibitemOpen
  \bibfield  {author} {\bibinfo {author} {\bibfnamefont {K.}~\bibnamefont
  {Stannigel}}, \bibinfo {author} {\bibfnamefont {P.}~\bibnamefont {Komar}},
  \bibinfo {author} {\bibfnamefont {S.~J.~M.}\ \bibnamefont {Habraken}},
  \bibinfo {author} {\bibfnamefont {S.~D.}\ \bibnamefont {Bennett}}, \bibinfo
  {author} {\bibfnamefont {M.~D.}\ \bibnamefont {Lukin}}, \bibinfo {author}
  {\bibfnamefont {P.}~\bibnamefont {Zoller}}, \ and\ \bibinfo {author}
  {\bibfnamefont {P.}~\bibnamefont {Rabl}},\ }\bibfield  {title} {\enquote
  {\bibinfo {title} {Optomechanical quantum information processing with photons
  and phonons},}\ }\href {\doibase 10.1103/PhysRevLett.109.013603} {\bibfield
  {journal} {\bibinfo  {journal} {Phys. Rev. Lett.}\ }\textbf {\bibinfo
  {volume} {109}},\ \bibinfo {pages} {013603} (\bibinfo {year}
  {2012})}\BibitemShut {NoStop}%
\bibitem [{\citenamefont {Pirkkalainen}\ \emph {et~al.}(2015)\citenamefont
  {Pirkkalainen}, \citenamefont {Cho}, \citenamefont {Massel}, \citenamefont
  {Tuorila}, \citenamefont {Heikkil{\"a}}, \citenamefont {Hakonen},\ and\
  \citenamefont {Sillanp{\"a}{\"a}}}]{Pirkkalainen2015}%
  \BibitemOpen
  \bibfield  {author} {\bibinfo {author} {\bibfnamefont {J.-M.}\ \bibnamefont
  {Pirkkalainen}}, \bibinfo {author} {\bibfnamefont {S.~U.}\ \bibnamefont
  {Cho}}, \bibinfo {author} {\bibfnamefont {F.}~\bibnamefont {Massel}},
  \bibinfo {author} {\bibfnamefont {J.}~\bibnamefont {Tuorila}}, \bibinfo
  {author} {\bibfnamefont {T.~T.}\ \bibnamefont {Heikkil{\"a}}}, \bibinfo
  {author} {\bibfnamefont {P.~J.}\ \bibnamefont {Hakonen}}, \ and\ \bibinfo
  {author} {\bibfnamefont {M.~A.}\ \bibnamefont {Sillanp{\"a}{\"a}}},\
  }\bibfield  {title} {\enquote {\bibinfo {title} {Cavity optomechanics
  mediated by a quantum two-level system},}\ }\href
  {https://www.nature.com/articles/ncomms7981} {\bibfield  {journal} {\bibinfo
  {journal} {Nat. Commun.}\ }\textbf {\bibinfo {volume} {6}},\ \bibinfo {pages}
  {6981} (\bibinfo {year} {2015})}\BibitemShut {NoStop}%
\bibitem [{\citenamefont {Scully}\ and\ \citenamefont
  {Zubairy}(1997)}]{Scully1997}%
  \BibitemOpen
  \bibfield  {author} {\bibinfo {author} {\bibfnamefont {M.~O.}\ \bibnamefont
  {Scully}}\ and\ \bibinfo {author} {\bibfnamefont {M.~S.}\ \bibnamefont
  {Zubairy}},\ }\href@noop {} {\emph {\bibinfo {title} {Quantum Optics}}}\
  (\bibinfo  {publisher} {Cambridge University Press},\ \bibinfo {year}
  {1997})\BibitemShut {NoStop}%
\bibitem [{\citenamefont {Cleland}\ and\ \citenamefont
  {Geller}(2004)}]{Cleland2004}%
  \BibitemOpen
  \bibfield  {author} {\bibinfo {author} {\bibfnamefont {A.~N.}\ \bibnamefont
  {Cleland}}\ and\ \bibinfo {author} {\bibfnamefont {M.~R.}\ \bibnamefont
  {Geller}},\ }\bibfield  {title} {\enquote {\bibinfo {title} {Superconducting
  qubit storage and entanglement with nanomechanical resonators},}\ }\href
  {\doibase 10.1103/PhysRevLett.93.070501} {\bibfield  {journal} {\bibinfo
  {journal} {Phys. Rev. Lett.}\ }\textbf {\bibinfo {volume} {93}},\ \bibinfo
  {pages} {070501} (\bibinfo {year} {2004})}\BibitemShut {NoStop}%
\bibitem [{\citenamefont {Zueco}\ \emph {et~al.}(2009)\citenamefont {Zueco},
  \citenamefont {Reuther}, \citenamefont {Kohler},\ and\ \citenamefont
  {H\"anggi}}]{Zueco2009}%
  \BibitemOpen
  \bibfield  {author} {\bibinfo {author} {\bibfnamefont {D.}~\bibnamefont
  {Zueco}}, \bibinfo {author} {\bibfnamefont {G.~M.}\ \bibnamefont {Reuther}},
  \bibinfo {author} {\bibfnamefont {S.}~\bibnamefont {Kohler}}, \ and\ \bibinfo
  {author} {\bibfnamefont {P.}~\bibnamefont {H\"anggi}},\ }\bibfield  {title}
  {\enquote {\bibinfo {title} {Qubit-oscillator dynamics in the dispersive
  regime: Analytical theory beyond the rotating-wave approximation},}\ }\href
  {\doibase 10.1103/PhysRevA.80.033846} {\bibfield  {journal} {\bibinfo
  {journal} {Phys. Rev. A}\ }\textbf {\bibinfo {volume} {80}},\ \bibinfo
  {pages} {033846} (\bibinfo {year} {2009})}\BibitemShut {NoStop}%
\bibitem [{\citenamefont {O'Connell}\ and\ \citenamefont
  {Cleland}(2014)}]{OConnell2014}%
  \BibitemOpen
  \bibfield  {author} {\bibinfo {author} {\bibfnamefont {A.}~\bibnamefont
  {O'Connell}}\ and\ \bibinfo {author} {\bibfnamefont {A.~N.}\ \bibnamefont
  {Cleland}},\ }\bibfield  {title} {\enquote {\bibinfo {title}
  {Microwave-frequency mechanical resonators operated in the quantum limit},}\
  }in\ \href {\doibase 10.1007/978-3-642-55312-7_12} {\emph {\bibinfo
  {booktitle} {Cavity Optomechanics}}}\ (\bibinfo  {publisher} {Springer},\
  \bibinfo {year} {2014})\ pp.\ \bibinfo {pages} {253--281}\BibitemShut
  {NoStop}%
\end{thebibliography}%

\end{document}